\documentclass[twocolumn,preprintnumbers,amsmath,amssymb]{revtex4} 
\usepackage{epsfig}
\usepackage{ae}
\usepackage{bm} 
\usepackage{color}
\usepackage{amsmath}
\usepackage{amssymb}
\usepackage{amsfonts}
\usepackage{mathbbol}
\usepackage{graphicx}
\usepackage{slashed}
\usepackage{comment}

\begin{document}
\title{Integrating out the Dirac sea in the Walecka model}
\author{Felix Karbstein\footnote{Electronic address:  felix@theorie3.physik.uni-erlangen.de}}
\affiliation{Institut f\"ur Theoretische Physik III,
Universit\"at Erlangen-N\"urnberg, D-91058 Erlangen, Germany}
\date{\today}

\begin{abstract}
We derive a purely fermionic no-sea effective theory, featuring positive-energy states only for the Walecka model.
In contrast to the so-called mean-field theory approach with the no-sea approximation, where the Dirac sea is simply omitted from the outset, we turn to the relativistic Hartree approximation and explicitly construct a no-sea effective theory from the underlying quantum field theory.
Several results obtained within these two approaches are confronted with each other. This sheds new light on the reliability of the mean-field theory with the no-sea approximation as well as the role of the Dirac sea.
Restricting to 1+1 dimensions, we obtain new analytical insights into nonuniform nuclear matter.
\end{abstract}

\maketitle

\section{Introduction}

The Walecka model, also known as $(\sigma,\,\omega)$ model or quantum hadrodynamics I (QHD-I), is a simple relativistic quantum field theory (QFT), mimicking the main features of the nucleon-nucleon interaction:
Forces between nucleons (spinor field $\psi$) are mediated by the exchange of mesons. As far as isospin symmetric matter (equal number of protons and neutrons) is concerned, the dominant one meson exchanges come from the scalar sigma ($\sigma$) meson and vector omega ($\omega$) meson.
The Walecka model defined by the following Lagrangian \cite{Walecka:1974qa,Boguta:1977xi,Serot:1997xg}
\begin{align}
  \mathcal{L}=&\;\bar\psi\left(\gamma_{\mu}({\rm i}\partial^{\mu}-\tilde{g}_{\omega}\omega^{\mu})-(\tilde{m}-\tilde{g}_{\sigma}\sigma)\right)\psi \nonumber\\
  &{}+\frac{1}{2}\left(\partial_{\mu}\sigma\partial^{\mu}\sigma-\tilde{m}_{\sigma}^2\sigma^2\right)+\frac{1}{3!}\tilde{c}_3\sigma^3+\frac{1}{4!}\tilde{c}_4\sigma^4  \nonumber\\
  &{}-\frac{1}{4}F_{\mu\nu}F^{\mu\nu}+\frac{1}{2}\tilde{m}_{\omega}^2\omega_{\mu}\omega^{\mu}\,,
  \label{Lag3+1}
\end{align}
where $F^{\mu\nu}=\partial^{\mu}\omega^{\nu}-\partial^{\nu}\omega^{\mu}$ is the kinetic term associated with the vector meson field $\omega^{\mu}$. The nucleon field is assumed to have $N$ flavors, i.e., $\bar\psi\psi=\sum_{k=1}^N\bar\psi_k\psi_k$ etc. In a phenomenological context one mostly works with $N=2$, where the two indices denote the proton and neutron fields, respectively.
From the viewpoint of {\it renormalized perturbation theory}, the cubic and quartic scalar self-interactions have to be included in the Lagrangian, to make the model renormalizable in 3+1 dimensions. Only then the respective counter terms $\sim\sigma^3$ and $\sim\sigma^4$ arise systematically, when rescaling the bare fields in Eq.~(\ref{Lag3+1}) and splitting the Lagrangian into a physical and a counterterm part (see Sec.~\ref{2a}) in the usual way (cf. e.g. Ref.~\cite{Peskin:1995ev}).

As the physical coupling constants are assumed to be large in this model, perturbation theory is inapplicable. However, mean-field theory (MFT) with the no-sea approximation, the relativistic Hartree approximation (RHA), and the Hartree-Fock (HF) approximation are known to yield physically sensible results.
The basic idea behind these approximations is to reduce the underlying interacting QFT to the problem of a single particle moving within the effective potential generated by all other particles.
Here we turn to the simplest approach of this type, the relativistic Hartree approximation.
In this context, one of the main problems in any practical calculation is posed by the question of how to account for the infinite number of negative-energy states constituting the Dirac sea.
Due to this difficulty, early mean-field type calculations in attempts to base nuclear physics on field theoretical models like the Walecka model, were typically done without the sea.
Similar ansätze were also considered for the Gross-Neveu (GN) model family in the realm of strong interaction physics.

Recently a new method to derive a no-sea effective theory (NSET) with positive-energy states only has been introduced and validated within the framework of the GN model family in 1+1 dimensions, where various exact analytical solutions are known. Moreover, new analytical insights were obtained \cite{Karbstein:2007be}. In the course of the NSET approach the Dirac sea is integrated out explicitly. Most strikingly this gives rise to new, Dirac sea-induced interactions (between positive-energy states) not present in the original Lagrangian. The coupling constants and interaction terms of the resulting effective theory are well defined and finite. The NSET is free of any UV divergences.
As the derivation of the NSET involves only standard Feynman diagrams, the approach is quite general and can be easily extended to different QFTs amenable to a mean-field type approximation.
In the present work, we apply this method to the Walecka model in 1+1 and 3+1 dimensions.
While the Walecka model was originally formulated in 3+1 dimensions \cite{Walecka:1974qa}, there are also studies in 1+1 dimensions, where calculational efforts are less and even some exact solutions are known \cite{Ferree:1993zz}.
Our main interest is in the model in 3+1 dimensions.
However, as from the viewpoint of the NSET, the models in 1+1 and 3+1 dimensions can be treated in a similar way, and on the other hand, fully analytical calculations are feasible in 1+1 dimensions, we find it worthwhile to consider also the model in 1+1 dimensions here.

This article is organized as follows. In Sec.~\ref{2} we construct the NSET for the Walecka model. In the course of this, renormalization conditions are specified. As it is conventionally done for the Walecka model, we utilize counterterm renormalization.
In Sec.~\ref{3} a test of the NSET is performed by restricting to uniform nuclear matter.
Section~\ref{3a} focuses on nonuniform nuclear matter in the model in 1+1 dimensions. Following the NSET approach, explicit analytical calculations can be performed here.
Our main focus will be on the question of identifying an expansion parameter that allows us to perform consistent and systematic approximations of the full problem.
We intend to provide some first insights in the abilities of the NSET approach in the treatment of spatially nonuniform systems.
Section~\ref{4} discusses some general observations and results obtained for the Walecka model in the NSET approach. In particular we contrast our findings with those obtained within MFT, were the Dirac sea is simply ignored from the outset. Several differences are stressed. We end with a short summary and conclusions in Sec.~\ref{5}.  

\section{Construction of no-sea effective theory} \label{2}
\subsection{Prerequisites}\label{2a}

\begin{figure}
\begin{center}
\epsfig{file=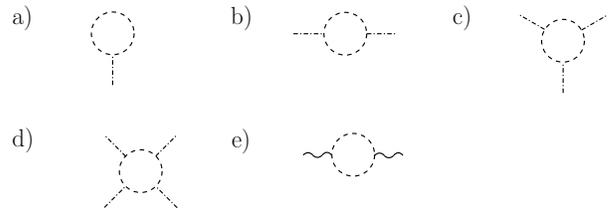,width=8.0cm,angle=0}
\end{center}
\caption{Divergent primitive Feynman diagrams. In 3+1 dimensions, diagrams a)-e) diverge. In 1+1 dimensions, diagrams a), b) and e) diverge, while c) and d) are finite. Dashed line: free fermion propagator; dash-dotted line: scalar propagator; curly line: vector propagator.}
\label{divdiags}
\end{figure}

In the course of counterterm renormalization, the original Lagrangian~(\ref{Lag3+1}) is written as $\mathcal{L}=\mathcal{L}'+\mathcal{L}_{CT}$,
with
\begin{align}
  \mathcal{L}'=&\;\bar\psi\left(\gamma_{\mu}({\rm i}\partial^{\mu}-{g}_{\omega}\omega^{\mu})-({m}-{g}_{\sigma}\sigma)\right)\psi \nonumber\\ &{}+\frac{1}{2}\left(\partial_{\mu}\sigma\partial^{\mu}\sigma-{m}_{\sigma}^2\sigma^2\right)+\frac{1}{3!}{c}_3\sigma^3+\frac{1}{4!}{c}_4\sigma^4 \nonumber\\
  &{}-\frac{1}{4}F_{\mu\nu}F^{\mu\nu}+\frac{1}{2}{m}_{\omega}^2\omega_{\mu}\omega^{\mu}
  \label{Lag3+1r}
\end{align}
and the counterterm Lagrangian
\begin{align}
  \mathcal{L}_{CT}=&-\delta_m\bar\psi\psi+\frac{1}{2!}\alpha_2\sigma^2+\frac{1}{3!}\alpha_3\sigma^3+\frac{1}{4!}\alpha_4\sigma^4 \nonumber\\
  &{}+\frac{1}{2}\alpha_{\sigma}\partial_{\mu}\sigma\partial^{\mu}\sigma+\frac{1}{4}\alpha_{\omega}F_{\mu\nu}F^{\mu\nu}\,.
  \label{LCT}
\end{align}
While the tilded parameters and the fields in Eq.~(\ref{Lag3+1}) are bare and unphysical, in Eq.~(\ref{Lag3+1r}) the analogous (untilded) parameters are considered as physical and measurable.
In this sense, the counterterm Lagrangian is not added. The original Lagrangian is rather split into two parts. $\mathcal{L}'$ is the physical part; $\mathcal{L}_{CT}$ contains the infinite but unobservable shifts between the bare and the physical parameters. As their choice is scale dependent, the physical parameters have to be specified by renormalization conditions, stating under which circumstances they have to be measured. Subsequently, the counterterms are tuned such that theory and measurement agree under these conditions.
Note that the individual counterterms can be unambiguously assigned to the (divergent) primitive diagrams depicted in Fig.~\ref{divdiags}.
Choosing the counterterms so as to cancel the divergences, the theory is rendered finite.
In result, a divergent primitive Feynman diagrams never occurs on its own, but always in its renormalized, finite form.

In order to remove the divergences in Figs.~\ref{divdiags}c,d in 3+1 dimensions, the scalar interactions $\sim\sigma^3$ and $\sim\sigma^4$ in the original Lagrangian~(\ref{Lag3+1}) are mandatory. Without incorporating them, the associated counterterms proportional to $\alpha_3$ and $\alpha_4$ would not be generated, when rescaling the fields in Eq.~(\ref{Lag3+1}) to their renormalized counterparts in Eq.~(\ref{Lag3+1r}), and the respective divergences could not be cured.
We will however assume (cf. Ref.~\cite{Serot:1984ey}) that the physical couplings, cubic and quartic in the scalar field, vanish in the vacuum at zero external momentum transfer, i.e., $c_3=c_4=0$. This choice actually corresponds to two renormalization conditions, fixing the explicit values of $\alpha_3$ and $\alpha_4$ (see below).

In 1+1 dimensions, the diagrams depicted in Figs.~\ref{divdiags}c,d are finite. Hence, the counterterm contributions $\sim \sigma^3$ and $\sim \sigma^4$ are not needed.
Working in 1+1 dimensions, we therefore set $\tilde{c}_3=\tilde{c}_4=0$ from the beginning.
Consequently, the respective counterterm contributions are not generated and $\alpha_3=\alpha_4=0$.

In the following, the divergent fermion loop diagrams are treated in dimensional regularization. Our conventions are as follows
\begin{equation}
\int{\rm d}^dp\,\frac{1}{(p^2+2pq-m^2)^n}=\frac{\frac{\Gamma\left(n-\frac{d}{2}\right)}{\Gamma(n)}(-1)^n{\rm i}\pi^{d/2}}{\left(m^2+q^2\right)^{n-d/2}}\,,
\label{dint}
\end{equation}
where $d=D+1-\epsilon$, with $\epsilon\to0$ in $D+1$ space-time dimensions.

While we choose and determine the values of the various counterterms as in the original work of Chin \cite{Chin:1977iz}, we consider it as helpful to shortly repeat the basic ideas of their determination in this context. In particular as our reasoning is somewhat different here and tailored to the formulation of the NSET. Instead of assembling the counterterms in a separate counterterm Lagrangian (cf. Eq.~(\ref{LCT})) and just resorting to them, when calculating observables, the counterterms are an integral ingredient of the NSET Lagrangian, thereby rendering the NSET free of any UV divergences.

\begin{figure}
	\begin{center}
	\epsfig{file=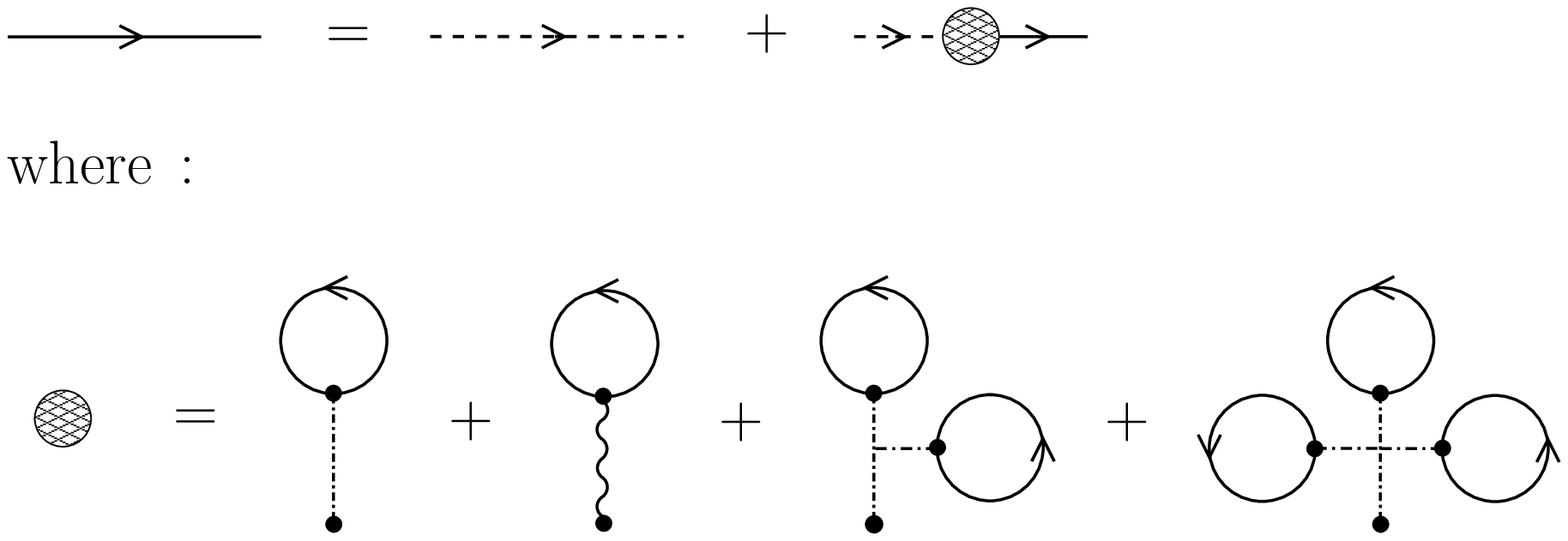,width=8.2cm,angle=0}
	\end{center}
	\caption{Dyson equation for fermion propagator in RHA. Obviously the last two terms in the second line do not arise in 1+1 dimensions. Dashed line: free propagator; solid line: dressed propagator; shaded circle: self-energy; dash-dotted line: scalar propagator; curly line: vector propagator.}
	\label{hartree}
\end{figure}

\begin{figure}
\begin{center}
\epsfig{file=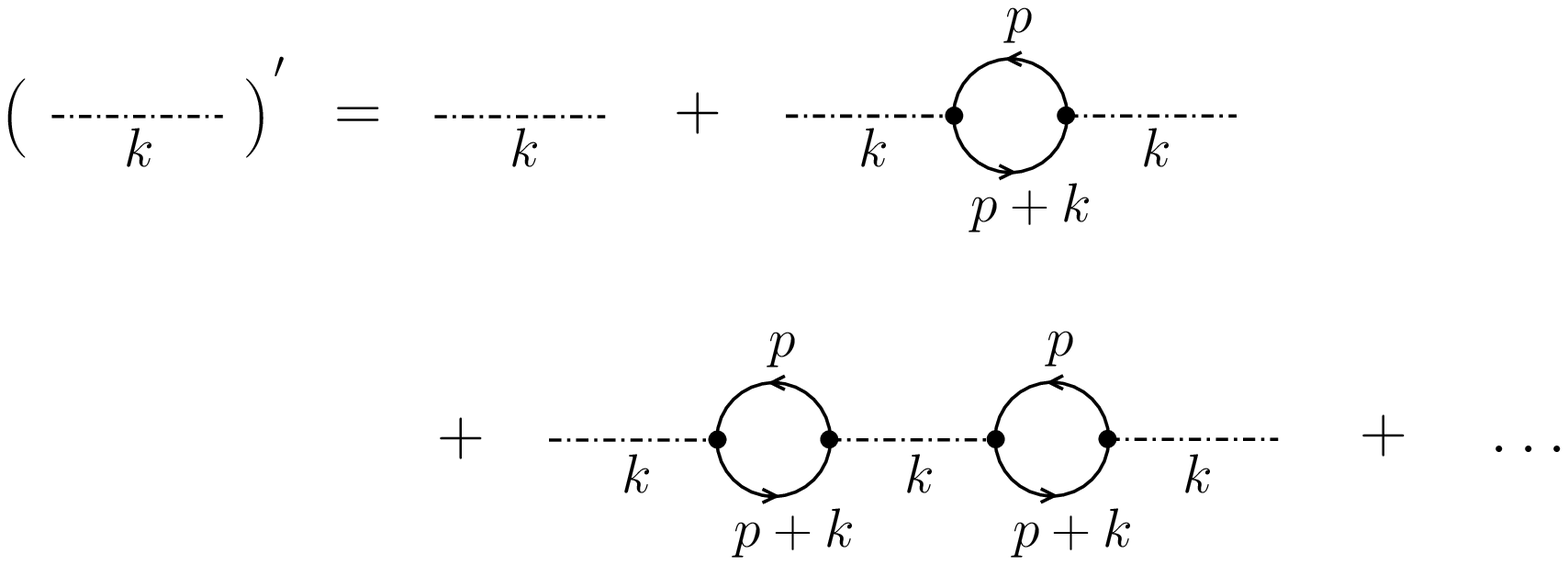,width=8.2cm,angle=0}
\end{center}
\caption{Resummation of scalar vacuum polarization graphs coupled via $\sigma$ propagators into a dressed scalar field propagator.}
\label{props}
\end{figure}

We first determine the nucleon self-energy in the vacuum. The vacuum is assumed to be translationally invariant.
In general the full fermion self-energy $\Sigma_H$ (cf. Fig.~\ref{hartree}) has scalar ($S$) as well as vector ($V_{\mu}$) contributions
\begin{equation}
 \Sigma_H=(S-m)-\gamma^{\mu}V_{\mu}\,.
\label{self-energy}
\end{equation}
However the vector contribution can be shown to vanish in the vacuum, resulting in a purely scalar contribution $\left.\Sigma_H\right|_{\rm vac}=M-m$. Here we introduced the physical fermion mass in the vacuum $M=S|_{\rm vac}$.
A common choice \cite{Serot:1984ey,Chin:1977iz} is
\begin{equation}
\left.\Sigma_H\right|_{\rm vac}=0\quad\leftrightarrow\quad M\equiv m\,,
\label{M=m}
\end{equation}
i.e., the mass $m$ occurring in the Lagrangian ${\cal L}={\cal L}'+{\cal L}_{CT}$ already corresponds to the physical mass in the vacuum.
In this sense, all possible loop corrections are already included and $\Sigma_H$ vanishes in the vacuum.
It is obvious from Fig.~\ref{hartree}, that the renormalization condition Eq.~(\ref{M=m}) is fulfilled by fixing $\delta_m$ such that the single primitive Feynman diagram Fig.~\ref{divdiags}a) vanishes at zero external momentum transfer
\begin{equation}
\delta_{m}=-N\frac{g_{\sigma}^2}{m_{\sigma}^2}\int\frac{{\rm d}^dp}{(2\pi)^d}{\rm Tr}\left\{\frac{\rm i}{\slashed{p}-m+{\rm i}\eta}\right\}\,.
\label{deltam}
\end{equation}
The values for the remaining counterterms are determined in the next step. Therefore we turn to the construction of dressed meson field propagators in the vacuum. In the presence of the cubic and quartic couplings of the scalar field this is rather complicated, but significantly simpler without them.
Let us for the moment turn to the Walecka model in 1+1 dimensions where $c_3=c_4=0$ and also $\alpha_3=\alpha_4=0$. For this choice, the Lagrangian is quadratic in the scalar as well as the vector field. Hence, both fields can be integrated out exactly, yielding a purely fermionic theory. We directly turn to the renormalized version, i.e., integrate out the renormalized meson fields. This yields
\begin{align}
  \mathcal{L}=&\;\bar\psi\left({\rm i}\slashed{\partial}-m-\delta_m\right)\psi \nonumber\\
  &{}+\frac{g_{\sigma}^2}{2}\bar\psi\psi\frac{1}{(1+\alpha_{\sigma})\square+m_{\sigma}^2-\alpha_2}\bar\psi\psi \nonumber\\ &{}+\frac{g_{\omega}^2}{2}\bar\psi\gamma^{\nu}\psi\frac{1}{(1+\alpha_{\omega})\square+m_{\omega}^2} \nonumber\\
  &\hspace*{16mm}{}\times\left(-\delta^{\ \mu}_{\nu}-\frac{(1+\alpha_{\omega})\partial_{\nu}\partial^{\mu}}{m_{\omega}^2}\right)\bar\psi\gamma_{\mu}\psi\,.
  \label{Lag1+1b}
\end{align} 
From Eq.~(\ref{Lag1+1b}) one can easily read off the (momentum space) propagators, incorporating the appropriate counterterms, for the $\sigma$ and $\omega^{\mu}$ fields. They are given by
\begin{equation}
{\rm i}\Delta(p)=\frac{\rm i}{(1+\alpha_{\sigma})p^2-m_{\sigma}^2+\alpha_2+{\rm i}\eta}
\label{prop1}
\end{equation}
and
\begin{align}
{\rm i}D^{\ \mu}_{\nu}(p)=&\;\frac{\rm i}{(1+\alpha_{\omega})p^2-m_{\omega}^2+{\rm i}\eta} \nonumber\\
&\times\left(-\delta^{\ \mu}_{\nu}+\frac{(1+\alpha_{\omega})p_{\nu}p^{\mu}}{m_{\omega}^2}\right)\,,
\label{prop2}
\end{align}
respectively. While the counterterms $\alpha_2$ and $\alpha_{\sigma}$ can be associated with ${\rm i}\Delta(p)$, $\alpha_{\omega}$ belongs to ${\rm i}D^{\ \mu}_{\nu}(p)$. The dressed versions of the propagators~(\ref{prop1}) and (\ref{prop2}) in the vacuum are now easily constructed. One simply has to account for all possible vacuum ($``-"$) fermion loop insertions in the respective meson propagator. To simplify the notation, the unrenormalized $``-"$ loop with two $1$ insertions and external momentum transfer $k$ is denoted by ${\rm i}\Sigma(k^2)$. We evaluate it by standard Feynman rules and expand the result in powers of the four-momentum $k$ flowing through the graph. Let us first consider the scalar field propagator. In RHA it emerges as depicted in Fig.~\ref{props}.
Summing up the geometric series, the corresponding analytical expression is
\begin{equation}
{\rm i}\Delta'(k)=\frac{\rm i}{(1+\alpha_{\sigma})k^2-m_{\sigma}^2+\alpha_2-g_{\sigma}^2\Sigma(k^2)+{\rm i}\eta}\,.
\label{dressedscalar}
\end{equation}
Note that both $\alpha_2$ as well as $\alpha_{\sigma}$ can be determined from $\Delta'(k)$.
In the following we turn to a particular choice of the counterterms.
Interpreting the parameter $m_{\sigma}^2$ to be the renormalized, physical mass of the $\sigma$ meson in the vacuum, we demand that the effective scalar propagator is of the form
\begin{equation}
 {\rm i}\Delta'(k)=\frac{\rm i}{k^2-m_{\sigma}^2+{\rm i}\eta}+{\cal O}(k^4)\,.
 \label{Deltastrichalg}
\end{equation}
In order to determine the effective vector field propagator, one has to invoke the same steps as for the effective scalar propagator.
We just have to replace the scalar couplings by vector couplings and the scalar propagators by vector propagators in Fig.~\ref{props}.
Due to the Lorentz indices, the determination of the effective vector propagator is somewhat more involved.
Performing the resummation in terms of a geometric series, a matrix in Lorentz indices has to be inverted.
To simplify the notation, the unrenormalized $``-"$ loop with $\gamma^{\mu}$ as well as $\gamma^{\nu}$ insertion at external momentum transfer $k$ is referred to as ${\rm i}\Pi^{\mu\nu}(k)$ in the following.
Extracting the tensor structure in the Lorentz indices 
\begin{equation}
{\rm i}\Pi^{\mu\nu}(k)={\rm i}(g^{\mu\nu}k^2-k^{\mu}k^{\nu})\Pi(k^2)\,,
\end{equation}
the vector propagator in the effective theory can be cast into the following algebraic form
\begin{align}
{\rm i}D'^{\ \mu}_{\nu}(k)=&\;\frac{\rm i}{k^2-m_{\omega}^2+{\rm i}\eta}\left[\frac{1}{1-C(k^2)k^2}\right. \nonumber\\
&{}\times\left.(-\delta_{\nu}^{\ \mu}+C(k^2)k_{\nu}k^{\mu})+\frac{k_{\nu}k^{\mu}}{m_{\omega}^2}\right]\,,
\label{Dstrichalg}
\end{align}
where we have introduced the momentum-dependent scalar function
\begin{equation}
C(k^2)=\frac{{\rm i}}{k^2-m_{\omega}^2+{\rm i}\eta}\left[g_{\omega}^2{\rm i}\Pi(k^2)+{\rm i}\alpha_{\omega}\right]\,.
\end{equation}
Note that for $C(k^2)\equiv0$, the vector field propagator of the original theory, Eq.~(\ref{prop2}), is recovered.
We fix $\alpha_{\omega}$ by implementing this choice.

\begin{figure}
\begin{center}
\epsfig{file=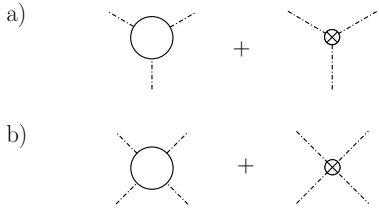,width=5cm,angle=0}
\end{center}
\caption{Diagrams giving rise to the physical interactions a) $\sim\sigma^3$ and b) $\sim\sigma^4$. The divergent fermion loops are renormalized by an adequate choice of the counterterms $\alpha_3$ and $\alpha_4$.}
\label{alpha34}
\end{figure}

Let us now reconsider the situation. To simplify the construction of the dressed meson propagators we turned to 1+1 dimensions and omitted the scalar interaction terms $\sim\sigma^3$ and $\sim\sigma^4$ for a moment.
However, it turns out that the expressions for the dressed meson propagators remain valid in 3+1 dimensions for zero physical couplings $c_3=c_4=0$. As noted above, this choice corresponds to two renormalization conditions fixing both $\alpha_3$ and $\alpha_4$.
The physical $\sigma^3$ and $\sigma^4$ interactions are depicted in Fig.~\ref{alpha34}.
We fix $\alpha_3$ and $\alpha_4$ in such a way that they exactly cancel the vacuum loops with three and four scalar insertions.
Hence, with the choice $c_3=c_4=0$ any additional contributions to the dressed meson propagators ${\rm i}\Delta'(k)$ and ${\rm i}D'^{\ \mu}_{\nu}(k)$ in the vacuum are ruled out by the mutual cancellation of the diagrams depicted in Figs.~\ref{divdiags}a) and \ref{divdiags}b), respectively.
Note that nonvanishing physical coefficients $c_3$ and $c_4$ would significantly complicate calculations, but are not required for a consistent model.
Finally all counterterm parameters have been fixed.
While $\delta_m$ and the counterterms multiplying powers of the meson fields ($\alpha_2\cdots\alpha_4$) are infinite, those multiplying derivatives of the meson fields ($\alpha_{\sigma}$, $\alpha_{\omega}$) are finite. The role of the latter is to ensure that the meson propagators have the expected momentum dependence.
We are now prepared to construct the no-sea effective theory with positive-energy states only.

\subsection{No-sea effective Lagrangian}

Let us now turn to the problem of finite fermion number. The RHA still has the same basic structure as in the vacuum, except that the nucleon self-energy is in general space-dependent and that the vector contribution in Fig.~\ref{hartree} no longer vanishes. The nucleon propagator gets an extra contribution from the positive-energy valence states.
We denote the vacuum and valence particle contributions by ``$-$" and ``$+$," respectively. 

\begin{figure}
\begin{center}
\epsfig{file=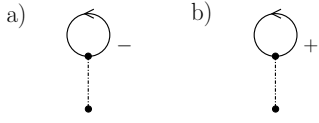,width=4.2cm,angle=0}
\end{center}
\caption{Decomposition of the one-loop contribution to the nucleon self-energy into a) negative and b) positive-energy contributions.}
\label{selfE1loop}
\end{figure}

\begin{figure}
\begin{center}
\epsfig{file=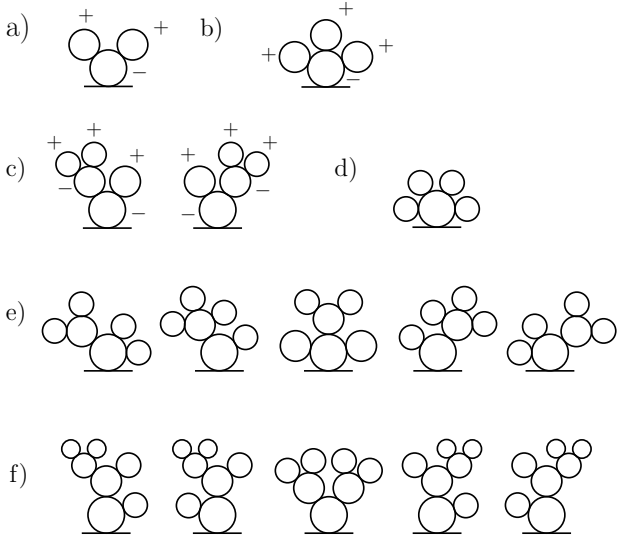,width=8.2cm,angle=0}
\end{center}
\caption{Contributions to the nucleon self-energy that are the seed for new interaction terms in the effective Lagrangian. For the sake of clarity, the ``$+$" and ``$-$" labels are only depicted in the first few diagrams and omitted in the remainder.}
\label{selfEdiags}
\end{figure}

Somewhat analogous to the decomposition of the nucleon Hartree propagator at finite Fermi momentum $k_f$, featuring a mass $M\equiv M(k_f)$ and accounting for a constant vector field $V^{\mu}\equiv V^{\mu}(k_f)$ \cite{Serot:1984ey}
\begin{align}
  {\rm i}G^H(p)&={\rm i}G_F^H(p)+{\rm i}G_D^H(p)\,, \nonumber\\
  {\rm i}G_F^H(p)&=\frac{\rm i}{\bar{\slashed{p}}-M+{\rm i}\eta}\,, \nonumber\\
  {\rm i}G_D^H(p)&=-\frac{\pi(\bar{\slashed{p}}+M)}{E(\bar{p})}\delta(\bar{p}^0-E(\bar{p}))\theta(k_f-|\bar{\bf{p}}|)\,,
\label{hartreepropf}
\end{align}
where $\bar{p}^{\mu}=p^{\mu}+V^{\mu}$ and $E(\bar{p})=\sqrt{\bar{\bf p}^2+M^2}$, we define
\begin{align}
  {\rm i}G^H(p)&={\rm i}G_-(p)+{\rm i}G_+(p)\,, \nonumber\\
  {\rm i}G_-(p)&=\frac{\rm i}{{\slashed{p}}-m+{\rm i}\eta}\,.
\label{hartreepropf}
\end{align}
Note that in contrast to ${\rm i}G_F^H(p)$, which has the form of the fermion Hartree propagator in the vacuum, - but contains the mass determined at finite Fermi momentum $M(k_f)$, ${\rm i}G_-(p)$ corresponds directly to the fermion Hartree propagator in the vacuum [mass $m=M(k_f=0)$)].
The second term ${\rm i}G_+^H(p)$ can be associated with the positive-energy contribution. Its explicit form is not important for the following discussion.
Hence, the one-loop diagram contributing to the self-energy for instance, then naturally splits up into the two pieces shown in Fig.~\ref{selfE1loop}.
In the effective theory with the Dirac sea integrated out, diagram Fig.~\ref{selfE1loop}a) is accounted for by the mass
term for the nucleon field
\begin{equation}
{\cal L}_{\rm eff}^{(1)}=-m\bar\psi\psi\,.
\end{equation}
(From now on the superscript on ${\cal L}_{\rm eff}$ refers to the power in the nucleon condensates $\bar\psi\psi$ and $\bar\psi\gamma^{\mu}\psi$ of the respective contribution to the NSET Lagrangian.) All the nucleon spinors appearing in the remainder of this article refer to the ``$+$" sector, hence we will omit the ``$+$" label. Diagram Fig.~\ref{selfE1loop}b) corresponds to the lowest-order no-sea RHA calculation with the interaction mediated by the original scalar propagator descending from Eq.~(\ref{Lag3+1}).
Obviously the dressed meson propagators can now be reinterpreted as the effective propagators mediating between nucleon ``$+$" fields. They already contain all possible ``$-$" loop insertions. Hence the four-fermion interactions of the effective Lagrangian can be written as
\begin{align}
 {\cal L}_{\rm eff}^{(2)}=\;&-\frac{g_{\sigma}^2}{2}(\bar\psi\psi)\Delta'(\bar\psi\psi) \nonumber\\
  &- \frac{g_{\omega}^2}{2}(\bar\psi\gamma^{\nu}\psi)D'^{\ \mu}_{\nu}(\bar\psi\gamma_{\mu}\psi)\,.
 \label{Leff2int}
\end{align}
The presence of the propagators in Eq.~(\ref{Leff2int}) shows that we are dealing with nonlocal effective interactions between positive-energy fermions. Here, however, our aim is to derive a local no-sea effective theory, valid in the vicinity of the fully occupied Dirac sea only.
We restrict ourselves to small external two-momentum transfer $k$ in the effective propagators Eqs.~(\ref{dressedscalar}) and (\ref{Dstrichalg}).
This will enable us to derive an ``almost local" effective Lagrangian well suited for analytical studies.
Let us therefore trade the effective propagators for effective, $k$-dependent couplings. 
As both the effective scalar field propagator and the effective vector field propagator describe massive fields, the resulting couplings are well defined in the infrared and contain no singularities.
We assume that the mass terms dominate the inverse effective meson propagators, i.e., $m_{\sigma}^2\gg k^2$ and $m_{\omega}^2\gg k^2$, respectively.
To keep the derivation of the NSET compact and general, we introduce coefficients $c_{ijk}$ in the following. As will become obvious below, the indices successively refer to the powers of $(\bar\psi\psi)$, $(\bar\psi\gamma^{\mu}\psi)$ and the derivative $\partial^{\mu}$ of the respective interaction terms in the NSET Lagrangian. Their explicit values in 1+1 and 3+1 dimensions are listed in (\ref{c1+1}) and (\ref{c3+1}).
While the effective scalar coupling is then given by
\begin{equation}
g_{\sigma,{\rm eff}}^2(k)=\frac{g_{\sigma}^2}{m_{\sigma}^2}\left\{1+\frac{1}{m_{\sigma}^2}\left[k^2+c_{204}k^4+\mathcal{O}(k^6)\right]\right\}\,,
\label{gseff}
\end{equation}
the effective vector coupling is determined by
\begin{align}
\left(g_{\omega,{\rm eff}}^2(k)\right)_{\nu}^{\ \mu}=&\;\frac{g_{\omega}^2}{m_{\omega}^2}\left\{-\delta_{\nu}^{\ \mu}-\frac{k^2\delta_{\nu}^{\ \mu}-k_{\nu}k^{\mu}}{m_{\omega}^2}\right.\nonumber\\
&{}\times\left.\phantom{\frac{1}{1}}\hspace*{-4mm}\left[1+c_{024}k^2\right]+\mathcal{O}(k^6)\right\}\,.
\label{gveff}
\end{align}
As a result, the nonlocal, effective interactions in Eq.~(\ref{Leff2int}) decompose into an infinite number of local interactions (involving increasing powers of derivatives in position space, as $k^{\mu}\to-{\rm i}\partial^{\mu}$).
We obtain
\begin{align}
\mathcal{L}_{\rm eff}^{(2,s)}=&\;\frac{g_{\sigma}^2}{2m_{\sigma}^2}\left(\bar\psi\psi\right)^2-\frac{g_{\sigma}^2}{2m_{\sigma}^4}\left(\square\bar\psi\psi\right)\left(\bar\psi\psi\right) \nonumber\\
&{}+\frac{g_{\sigma}^2}{2m_{\sigma}^4}c_{204}(\square^2\bar\psi\psi)(\bar\psi\psi)+\cdots
\label{L2s}
\end{align}
and
\begin{align}
\mathcal{L}_{\rm eff}^{(2,v)}=&-\frac{g_{\omega}^2}{2m_{\omega}^2}(\bar\psi\gamma_{\nu}\psi)(\bar\psi\gamma^{\nu}\psi)  \nonumber\\
&{}+\frac{g_{\omega}^2}{2m_{\omega}^4}\!\!\left[(\partial_{\nu}\bar\psi\gamma^{\nu}\psi)(\partial_{\mu}\bar\psi\gamma^{\mu}\psi)+(\square\bar\psi\gamma_{\nu}\psi)(\bar\psi\gamma^{\nu}\psi)\right] \nonumber\\
&{}-\frac{g_{\omega}^2}{2m_{\omega}^4}c_{024}\left[(\partial_{\nu}\bar\psi\gamma^{\nu}\psi)(\square\partial_{\mu}\bar\psi\gamma^{\mu}\psi)\right. \nonumber\\
&\hspace*{2.2cm}{}+\left.(\square^2\bar\psi\gamma_{\nu}\psi)(\bar\psi\gamma^{\nu}\psi)\right]+\cdots
\end{align}
for the scalar and vector contributions, respectively. The ellipses denote higher derivative terms.
Moreover, effective interactions involving higher powers in the nucleon condensates are generated in the process of integrating out the Dirac sea.
Note that, having derived the effective couplings~(\ref{gseff}) and (\ref{gveff}), the remaining steps in the construction of the no-sea effective theory for the Walecka model coincide with those invoked in the context of the GN model family \cite{Karbstein:2007be}. The only difference being that instead of scalar and pseudoscalar insertions, we now have to deal with scalar and vector, i.e., $1$ and $\gamma^{\mu}$ insertions in the ``$-$" loops.
In order to determine the effective interaction terms it is convenient to evaluate nucleon self-energy contributions and derive the effective interactions therefrom. The respective self-energy contributions are made up of internal ``$-$" loops with more than two insertions and outermost ``$+$" loops, coupled via the effective point couplings defined in Eqs.~(\ref{gseff}) and (\ref{gveff}). 
In this article we will consider terms in the effective Lagrangian that are $\sim$(nucleon condensates)$^5$ at most. Figure~\ref{selfEdiags} depicts the self-energy diagrams giving rise to effective interaction terms of the power $3$ [[Fig.~\ref{selfEdiags}a], $4$ [Fig.~\ref{selfEdiags}b) and \ref{selfEdiags}c)] and $5$ [Fig.~\ref{selfEdiags}d)-\ref{selfEdiags}f)] in the nucleon condensates.
To determine a given effective interaction term in the NSET Lagrangian, one has to sum up all topologically distinct diagrams with the respective number of outermost ``$+$" loops. Due to the infinite series making up the effective couplings, every of the diagrams in Fig.~\ref{selfEdiags} represents an infinite series of loop diagrams. Remember that the mass term already includes an infinite number of loops due to the self-consistency condition (see Fig.~\ref{hartree}). Moreover it can be shown, that all ``$-$" loops with at least one $\gamma^{\mu}$ insertion (and an arbitrary number of $1$ insertions) vanish at zero external momentum transfer.
In this article we do not explicitly derive all the interaction terms in the effective Lagrangian. We just illustrate the basic procedure for the effective interaction arising from Fig.~\ref{selfEdiags}a), evaluated with scalar interactions only. The underlying idea is an expansion under the loop integral in powers of the momentum transfer.
The analytical expression for the respective self-energy contribution is
\begin{align}
\delta\Sigma(k)=\;&-{\rm i}\int\frac{{\rm d}k_1}{2\pi}\frac{{\rm d}k_2}{2\pi}(2\pi)\delta(k-k_1-k_2) \nonumber\\
&{}\times\left({\rm i}g_{\sigma,{\rm eff}}^2(k_1+k_2)\right)\left({\rm i}g_{\sigma,{\rm eff}}^2(k_1)\right)\left({\rm i}g_{\sigma,{\rm eff}}^2(k_2)\right) \nonumber\\
&{}\times\frac{1}{2m}\frac{N}{(2\pi)^2}\left(k_1^2+k_1k_2+k_2^2+\mathcal{O}(k^4)\right) \nonumber\\
&{}\times\langle\bar\psi\psi\rangle_{k_1}\langle\bar\psi\psi\rangle_{k_2}\,,
\end{align}
where we treat both $k_1^{\mu}$ and $k_2^{\mu}$ as of ${\cal O}(k)$.
Making an ansatz for the interaction term in the effective Lagrangian
\begin{align}
\mathcal{L}_{\rm eff}^{(3,1)}=&\;\left(\frac{g_{\sigma}^2}{m_{\sigma}^2}\right)^3\left[c_{300}(\bar\psi\psi)^3\right. \nonumber\\
&\hspace*{1cm}{}+\left.c_{302}(\square\bar\psi\psi)(\bar\psi\psi)^2+\cdots\right]
\label{ansatz}
\end{align}
and deriving the corresponding self-energy contribution therefrom, the parameters $c_{300}$ and $c_{302}$ in Eq.~(\ref{ansatz}) can be inferred.
The other effective interactions are determined analogously. 
It turns out that there is just one further nonvanishing contribution from Fig.~\ref{selfEdiags}a, namely that arising from the ``$-$" loop with $\gamma^{\mu}$, $\gamma^{\nu}$, and $1$ insertion.
In this case, the interaction term in the effective Lagrangian is given by
\begin{align}
\mathcal{L}_{\rm eff}^{(3,2)}=\,&-\frac{g_{\sigma}^2}{m_{\sigma}^2}\left(\frac{g_{\omega}^2}{m_{\omega}^2}\right)^2c_{122}\,(\bar\psi\psi)\!\left[\partial^{\beta}(\bar\psi\gamma_{\alpha}\psi)\partial_{\beta}(\bar\psi\gamma^{\alpha}\psi)\right. \nonumber\\
&\hspace*{1cm}{}-\left.\partial^{\beta}(\bar\psi\gamma_{\alpha}\psi)\partial^{\alpha}(\bar\psi\gamma_{\beta}\psi)\right]{}+\cdots\,.
\end{align}
Note that due to the reason given above no new non-derivative interactions involving higher powers of the condensate $\bar\psi\gamma^{\mu}\psi$ are induced.
We do not continue in a fully systematic way beyond this level. However, in view of the applications we have in mind, we single out specific terms. Namely the non-derivative terms $\sim(\bar\psi\psi)^4$ and $\sim(\bar\psi\psi)^5$.
Collecting the different contributions, we finally arrive at the following NSET Lagrangian
\begin{widetext}
\begin{align}
\mathcal{L}_{\rm eff}=&\;\bar\psi({\rm i}\slashed{\partial}-m)\psi+\frac{g_{\sigma}^2}{2m_{\sigma}^2}(\bar\psi\psi)^2-\frac{g_{\sigma}^2}{2m_{\sigma}^4}(\square\bar\psi\psi)(\bar\psi\psi)+\frac{g_{\sigma}^2}{2m_{\sigma}^4}c_{204}(\square^2\bar\psi\psi)(\bar\psi\psi)-\frac{g_{\omega}^2}{2m_{\omega}^2}(\bar\psi\gamma_{\nu}\psi)(\bar\psi\gamma^{\nu}\psi)\nonumber\\
&{}+\frac{g_{\omega}^2}{2m_{\omega}^4}\left[(\partial_{\nu}\bar\psi\gamma^{\nu}\psi)(\partial_{\mu}\bar\psi\gamma^{\mu}\psi)+(\square\bar\psi\gamma_{\nu}\psi)(\bar\psi\gamma^{\nu}\psi)\right]-\frac{g_{\omega}^2}{2m_{\omega}^4}c_{024}\left[(\partial_{\nu}\bar\psi\gamma^{\nu}\psi)(\square\partial_{\mu}\bar\psi\gamma^{\mu}\psi)+(\square^2\bar\psi\gamma_{\nu}\psi)(\bar\psi\gamma^{\nu}\psi)\right] \nonumber\\
&{}-\frac{g_{\sigma}^2}{m_{\sigma}^2}\left(\frac{g_{\omega}^2}{m_{\omega}^2}\right)^2c_{122}
(\bar\psi\psi)\left[\partial^{\beta}\left(\bar\psi\gamma_{\alpha}\psi\right)\partial_{\beta}\left(\bar\psi\gamma^{\alpha}\psi\right)-\partial^{\beta}\left(\bar\psi\gamma_{\alpha}\psi\right)\partial^{\alpha}\left(\bar\psi\gamma_{\beta}\psi\right)\right]+\left(\frac{g_{\sigma}^2}{m_{\sigma}^2}\right)^3c_{300}(\bar\psi\psi)^3 \nonumber\\
&{}+\left(\frac{g_{\sigma}^2}{m_{\sigma}^2}\right)^3c_{302}(\square\bar\psi\psi)(\bar\psi\psi)^2+\left(\frac{g_{\sigma}^2}{m_{\sigma}^2}\right)^4c_{400}(\bar\psi\psi)^4+\left(\frac{g_{\sigma}^2}{m_{\sigma}^2}\right)^5c_{500}(\bar\psi\psi)^5{}+\cdots\,.
\label{effthind+1}
\end{align}
\end{widetext}
While the general structure of the effective Lagrangian is the same in 1+1 and 3+1 dimensions, the coefficients of the different interaction terms differ. We obtain
\begin{align}
c_{204}&=\frac{1}{m_{\sigma}^2}+\frac{Ng_{\sigma}^2}{120\pi m^4}\,,\quad c_{024}=\frac{1}{m_{\omega}^2}+\frac{Ng_{\omega}^2}{30\pi m^4}\,, \nonumber
\end{align}
\begin{align}
c_{122}&=\frac{N}{3\pi m^3}\,,\quad c_{300}=\frac{N}{6\pi m}\,,\quad c_{302}=-\frac{N}{24\pi m^3}\,,\nonumber\\
c_{400}&=\frac{N}{8\pi m^2}\left(\frac{1}{3}+\frac{Ng_{\sigma}^2}{\pi m_{\sigma}^2}\right)\,, \nonumber\\
c_{500}&=\frac{N}{20\pi m^3}\left[\frac{1}{3}+\frac{5Ng_{\sigma}^2}{\pi m_{\sigma}^2}\left(\frac{1}{3}+\frac{Ng_{\sigma}^2}{2\pi m_{\sigma}^2}\right)\right] \label{c1+1}
\end{align}
in 1+1 dimensions and
\begin{align}
c_{204}&=\frac{1}{m_{\sigma}^2}+\frac{Ng_{\sigma}^2}{80\pi^2m^2}\,,\quad c_{024}=\frac{1}{m_{\omega}^2}+\frac{Ng_{\omega}^2}{60\pi^2m^2}\,, \nonumber\\
c_{122}&=\frac{N}{12\pi^2m}\,,\quad c_{300}=0\,,\quad c_{302}=-\frac{N}{16\pi^2m}\,,\nonumber\\
c_{400}&=0\,,\quad c_{500}=-\frac{N}{40\pi^2m} \label{c3+1}
\end{align}
in 3+1 dimensions.
Let us emphasize again that the effective Lagrangian of the Walecka in 3+1 dimensions does not include non-derivative interactions cubic and quartic in the fermion bilinear $\bar\psi\psi$.
This is a consequence of our renormalization conditions (cf. Fig.~\ref{alpha34}). The determination of the $(\bar\psi\psi)^5$-interaction is therefore very simple in 3+1 dimensions. The only contributing diagram is Fig.~\ref{selfEdiags}d).

\section{Testing the no-sea effective theory} \label{3}

We propose to test the NSET Lagrangian for a low-density system of fermions with Fermi momentum $k_f$, assuming unbroken translational invariance. Only in this (possibly unphysical) case do we have the exact analytical solution to compare with.
If the condensates are assumed to be spatially uniform, only the non-derivative terms enter and Eq.~(\ref{effthind+1}) reduces to the following effective Lagrangian
\begin{align}
  \mathcal{L}_{\rm eff}'=&\;\bar\psi({\rm i}\slashed{\partial}-m)\psi+\frac{g_{\sigma}^2}{2m_{\sigma}^2}(\bar\psi\psi)^2-\frac{g_{\omega}^2}{2m_{\omega}^2}(\bar\psi\gamma_{\nu}\psi)(\bar\psi\gamma^{\nu}\psi) \nonumber\\
  &{}+\left(\frac{g_{\sigma}^2}{m_{\sigma}^2}\right)^3c_{300}(\bar\psi\psi)^3+\left(\frac{g_{\sigma}^2}{m_{\sigma}^2}\right)^4c_{400}(\bar\psi\psi)^4 \nonumber\\
  &{}+\left(\frac{g_{\sigma}^2}{m_{\sigma}^2}\right)^5c_{500}(\bar\psi\psi)^5{}+\cdots.
 \label{Lefftralainv}
\end{align}
The Euler-Lagrange equation then allows us to identify the self-energy $\Sigma_H$ via
\begin{equation}
  \frac{\partial}{\partial\bar\psi}\mathcal{L}_{\rm eff}'=\left({\rm i}\slashed{\partial}-\Sigma_H-m\right)\psi=0\,.
  \label{eq1-teft}
\end{equation}
Using the decomposition of the full self-energy into scalar and vector contributions, Eq.~(\ref{self-energy}), Eq.~(\ref{eq1-teft}) becomes
\begin{equation}
  \left({\rm i}\slashed{\partial}-M+\gamma^{\mu}V_{\mu}\right)\psi=0\,.
\end{equation}
The corresponding effective mass $M$ is then given by
\begin{align}
   M=&\;m-\frac{g_{\sigma}^2}{m_{\sigma}^2}\langle\bar\psi\psi\rangle-3\left(\frac{g_{\sigma}^2}{m_{\sigma}^2}\right)^3c_{300}\langle\bar\psi\psi\rangle^2 \label{Mtrala}\\
   &{}-4\left(\frac{g_{\sigma}^2}{m_{\sigma}^2}\right)^4\!c_{400}\langle\bar\psi\psi\rangle^3-5\left(\frac{g_{\sigma}^2}{m_{\sigma}^2}\right)^5\!c_{500}\langle\bar\psi\psi\rangle^4{}+\cdots\,.\nonumber
\end{align}
By construction the condensates that appear in Eq.~(\ref{Mtrala}) only refer to the positive-energy sector.
Using the familiar decomposition of the full fermion propagator (mass $M$)  at finite Fermi momentum $k_f$, into vacuum and valence particle contributions Eq.~(\ref{hartreepropf}),
the explicit expression for the desired condensate can be easily inferred
\begin{equation}
  \langle\bar\psi\psi\rangle=2N\int_0^{k_f}\frac{{\rm d}^{D}p}{(2\pi)^{D}}\,\frac{M}{\sqrt{{\bf p}^2+M^2}}\,,
\end{equation}
where $D$ refers to the space dimension.
This implies that the terms written explicitly in Eq.~(\ref{Mtrala}) should allow the determination of the effective mass valid up to ${\cal O}(k_f^{4})$ in 1+1 dimensions and up to ${\cal O}(k_f^{14})$ in 3+1 dimensions, respectively. The lowest-order term omitted in Eq.~(\ref{Mtrala}) is $\sim\langle\bar\psi\psi\rangle^5$.

Note that the underlying assumption is that $M$ can be expressed as a power series in $k_f$. That this is possible is not {\it a priori} clear \cite{Henning:1995sm}. However, here as well as in the case of the energy density, considered below, the exact RHA results are known and confirm this assumption within the Walecka model. 

For illustration let us quote the explicit expression of the low-density expansion of $M$ (expanding in $k_f$) in 3+1 dimensions. It is given by
\begin{align}
M=&\;m-\frac{c_{\sigma}}{3}(k_f)^3+\frac{c_{\sigma}}{10m^2}(k_f)^5-\frac{3c_{\sigma}}{56m^4}(k_f)^7 \nonumber\\
&{}+\frac{c_{\sigma}^2}{15m^3}(k_f)^8+\frac{5c_{\sigma}}{144m^6}(k_f)^9-\frac{16c_{\sigma}^2}{175m^5}(k_f)^{10} \nonumber\\
&{}+\frac{c_{\sigma}}{m^4}\left(\frac{c_{\sigma}^2}{30}-\frac{35}{1408m^4}\right)(k_f)^{11}{}+\cdots\,,
\label{Mtralaew}
\end{align}
where
\begin{equation}
  c_{\sigma}=\frac{Ng_{\sigma}^2}{\pi^2m_{\sigma}^2}\,.
  \label{csigma2}
\end{equation}
Here, $k_f$ is assumed to be small as compared to the mass scales $m$ and $m_{\sigma}$. More precisely, the expansion is in the dimensionless ratios $k_f/m$, $k_f/m_{\sigma}$ and $k_f/m_{\omega}$, respectively.
In contrast to the situation in the vacuum, the vector part of the self-energy also yields a finite contribution. As $\langle\bar\psi{\bm \gamma}\psi\rangle=0$ for a spatially uniform system, it is determined as follows
\begin{equation}
  V^{\nu}=-\frac{g_{\omega}^2}{m_{\omega}^2}\langle\bar\psi\gamma^{\nu}\psi\rangle=-\frac{g_{\omega}^2}{m_{\omega}^2}\rho_B\delta^{\nu}_0\,,
  \label{e1a}
\end{equation}
with
\begin{equation}
  \rho_B=\langle\psi^{\dag}\psi\rangle=2N\int_0^{k_f}\frac{{\rm d}^Dp}{(2\pi)^D}
  \,.
  \label{e2a}
\end{equation}
Let us now consider the exact solution.
Demanding translational invariance, we determine the fermion mass at finite $k_f$ in 3+1 dimensions \cite{Serot:1984ey}
\begin{align}
 M-m=&\;\frac{Ng_{\sigma}^2}{(2\pi)^2m_{\sigma}^2}\left\{-\frac{1}{\pi}\int_0^{k_f}{\rm d}^3p\frac{M}{\sqrt{p^2+M^2}} \right.\nonumber\\
 &{}+\left.2M^3\ln{\left(\frac{M}{m}\right)}-\frac{11}{3}M^3+6mM^2\right. \nonumber\\
 &{}-\left.3m^2M+\frac{2}{3}m^3\right\}\,.
 \label{Mm3+1}
\end{align}
The Taylor expansion of the exact result, Eq.~(\ref{Mm3+1}), reproduces Eq.~(\ref{Mtralaew}). The expression for the vector self-energy associated with Eq.~(\ref{Lefftralainv}), Eq.~(\ref{e1a}), agrees with the expression for the full vector self-energy, determined under the assumption of translational invariance.
This serves as a good test of our NSET in the special case of unbroken translational symmetry and confirms that the coefficients of the non-derivative terms in ${\cal L}_{\rm eff}$ have been evaluated correctly in 3+1 dimensions. In 1+1 dimensions the calculation is performed analogously.

No analytical bound state solutions are known for the Walecka model. Therefore we cannot test the derivative terms in the present case.
The analysis of localized bound states in 3+1 dimensions would presumably require numerical methods and is outside the scope of the present work. However, we can obtain some analytical insights in 1+1 dimensions. In the following section we demonstrate that the treatment of nonuniform nuclear matter within the NSET approach is indeed possible.  

\section{Application of no-sea effective theory in 1+1 dimensions} \label{3a}

So far, we have derived the no-sea effective theory for the Walecka model and tested it in a special case.
Here we use our effective Lagrangian~(\ref{effthind+1}) in 1+1 dimensions to study bound states of $n\leq N$ fermions.
Our main focus will be on the question of identifying an expansion parameter which allows us to perform consistent and systematic approximations to the full problem.

We start with the assumption that the leading-order interaction term in the effective theory is given by the two non-derivative four-fermion interactions
\begin{align}
\mathcal{L}_{\rm eff}=&\;\bar\psi({\rm i}\slashed{\partial}-m)\psi+\frac{g_{\sigma}^2}{2m_{\sigma}^2}(\bar\psi\psi)^2 \nonumber\\
&-\frac{g_{\omega}^2}{2m_{\omega}^2}(\bar\psi\gamma_{\nu}\psi)(\bar\psi\gamma^{\nu}\psi)\,.
\label{leegavlag}
\end{align}
In order to find a $n$-fermion bound-state solution associated with the effective Lagrangian~(\ref{leegavlag}), it is convenient to make use of an approach introduced by Lee {\it et al}. \cite{Lee:1975tx} to construct exact localized solutions of 1+1 dimensional field theories with four-fermion interactions.

By a slight generalization of their work to the simultaneous appearance of scalar and vector interactions, featuring two different coupling constants, this problem can be solved analytically. The derivation can be found in Appendix~\ref{app1}. This then enables us to organize and take into account higher order corrections to ${\cal L}_{\rm eff}$ in terms of a small parameter, at least in a certain regime.

It turns out that there is a solution only for
\begin{equation}
\frac{g_{\sigma}^2}{m_{\sigma}^2}>\frac{g_{\omega}^2}{m_{\omega}^2}\,.
\label{ineq}
\end{equation}
It remains to specify an expansion parameter. Therefore we invoke additional assumptions.
Let us set
\begin{equation}
\frac{1}{m_{\sigma}^2}=\frac{A}{\gamma m_0^2} \quad {\rm and} \quad \frac{1}{m_{\omega}^2}=\frac{B}{\gamma m_0^2}\,,
\label{defs}
\end{equation}
where $A$, $B$ and $\gamma$ are dimensionless parameters and $m_0$ is a unit mass.
Introducing three independent parameters $A$, $B$ and $\gamma$, the problem is overdetermined. We can set one of the parameters $A$ or $B$ equal to one.
This is done such that
\begin{equation}
{\rm max}\left(\frac{1}{m_{\sigma}^2},\frac{1}{m_{\omega}^2}\right)=\frac{1}{\gamma m_0^2}\,,
\label{bed}
\end{equation}
or, equivalently (cf. Eq.~(\ref{defs})),
\begin{equation}
{\rm max}\left(A,B\right)=1
\end{equation}
and
\begin{equation}
{\rm min}\left(A,B\right)=C\leq 1\,.
\end{equation}
Hence inequality~(\ref{ineq}) translates into
\begin{equation}
 C<\frac{g_{\sigma}^2}{g_{\omega}^2} \quad {\rm or} \quad C>\frac{g_{\omega}^2}{g_{\sigma}^2}\,,
\end{equation}
respectively.
Moreover, recall the conditions $m_{\sigma}^2\gg k^2$ and $m_{\omega}^2\gg k^2$, which were necessary in deriving a local effective theory.
Independent of Eq.~(\ref{bed}), they translate into $\gamma m_0^2\gg k^2$ and $\gamma m_0^2C^{-1} \gg k^2$. As $C\leq1$, both inequalities are simultaneously fulfilled if the single inequality $\gamma m_0^2\gg k^2$ is true. 
Here we want to use $\gamma^{-1}$ as our (dimensionless) expansion parameter. This means that we also have to guarantee $\gamma\gg 1$.
The specifications on the expansion parameter $\gamma^{-1}$ can thus be summarized by
\begin{equation}
\gamma m_0^2 \gg k^2 \quad {\rm and} \quad \gamma\gg 1\,.
\end{equation}
Incorporating Eq.~(\ref{defs}) and the appropriate specifications for $A$ and $B$ in the exact $n$-fermion bound-state solution of Eq.~(\ref{leegavlag}), given in Appendix~\ref{app1}, one finds
\begin{equation}
 \bar\psi\psi\sim\frac{1}{\gamma}\,, \quad \psi^{\dag}\psi\sim\frac{1}{\gamma}\,, \quad \bar\psi\gamma^1\psi=0\,,
 \label{+conds}
\end{equation}
the last equation being a direct consequence of our choice of the $\gamma$ matrices. Consequently, we have
\begin{equation}
 (\bar\psi\psi)^2\sim\frac{1}{\gamma^2}\,, \quad (\bar\psi\gamma^{\nu}\psi)^2\sim\frac{1}{\gamma^2}\,, \quad \partial_x\sim\frac{1}{\gamma}\,.
\end{equation}
We are now in a position to count $\gamma^{-1}$-powers of the various terms contributing to the full effective Lagrangian.
In accordance with our assumption, interaction terms featuring higher powers of the fermion bilinears are of higher order in $\gamma^{-1}$ and therefore subleading as compared to Eq.~(\ref{leegavlag}). Let us, for example, concentrate on the Lagrangian valid up to $\mathcal{O}(\gamma^{-7})$
\begin{align}
 \mathcal{L}_{\rm eff}=&\;\bar\psi({\rm i}\slashed{\partial}-m)\psi+\frac{g_{\sigma}^2}{2m_{\sigma}^2}(\bar\psi\psi)^2-\frac{g_{\omega}^2}{2m_{\omega}^2}(\bar\psi\gamma_{\nu}\psi)(\bar\psi\gamma^{\nu}\psi) \nonumber\\
 &{}+\frac{N}{6\pi m}\left(\frac{g_{\sigma}^2}{m_{\sigma}^2}\right)^3(\bar\psi\psi)^3-\frac{1}{2}\frac{g_{\sigma}^2}{m_{\sigma}^4}(\square\bar\psi\psi)(\bar\psi\psi) \nonumber\\
 &{}+\frac{1}{2}\frac{g_{\omega}^2}{m_{\omega}^4}(\bar\psi\gamma_{\nu}\psi)(\square\bar\psi\gamma^{\nu}\psi)\,.
 \label{Leffgamma6}
\end{align}
The fermion bound-state mass $M_n$ computed from this Lagrangian should be valid up to $\mathcal{O}(\gamma^{-6})$. As $\int{\rm d}x\sim\gamma$, the validity of the series expansion of $M_n$ is reduced by one power of $\gamma^{-1}$ as compared to the series expansion of ${\cal L}_{\rm eff}$.
In order to determine $M_n$, we follow the procedure developed above in the context of the NJL model and use a series expansion in $\gamma^{-1}$.
To do this, we turn to the equation of motion derived from Eq.~(\ref{Leffgamma6}), written in the following form
\begin{equation}
 \left(-\gamma_5{\rm i}\partial_x+\gamma^0S(x)-V_0(x)\right)\psi_{\alpha}=E_{\alpha}\psi_{\alpha}\,,
 \label{eqmotalpha}
\end{equation}
with
\begin{equation}
S(x)=m+s-\frac{N}{2\pi m}\frac{g_{\sigma}^2}{m_{\sigma}^2}\,s^2+\frac{1}{m_{\sigma}^2}\,\partial_x^2\,s
\label{S}
\end{equation}
and
\begin{equation}
V_0(x)=\rho_0+\frac{1}{m_{\omega}^2}\,\partial_x^2\,\rho_0
\end{equation}
expressed self-consistently through the condensates
\begin{equation}
s(x)=-\frac{g_{\sigma}^2}{m_{\sigma}^2}\langle\bar\psi\psi\rangle \quad {\rm and} \quad \rho_0(x)=-\frac{g_{\omega}^2}{m_{\omega}^2}\langle\psi^{\dag}\psi\rangle\,.
\end{equation}
Here we used that $\langle\bar\psi\gamma^1\psi\rangle$ vanishes [cf. Eq.~(\ref{+conds})]. We solve Eq.~(\ref{eqmotalpha}) for a single energy level $\alpha\equiv0$, assuming $n\leq N$.
If $Ag_{\sigma}^2-Bg_{\omega}^2>0$, the bound-state mass is then given by
\begin{align}
 \!\frac{M_n}{nm}=&\;1-c^2\left\{\frac{n^2}{24}\frac{1}{\gamma^2} -n^4c\frac{9Ag_{\sigma}^2-Bg_{\omega}^2}{1920\,m_0^2}\frac{1}{\gamma^4}\right. \label{MBexp}\\
 &{}+\left.\frac{n^4}{180}\left[\frac{N}{\pi}\!\left(\frac{Ag_{\sigma}^2}{m_0^2}\right)^3-3m^2c\frac{A^2g_{\sigma}^2-B^2g_{\omega}^2}{m_0^4}\right]\frac{1}{\gamma^5}
 \right\}\,,\nonumber 
\end{align}
where we defined $c=(Ag_{\sigma}^2-Bg_{\omega}^2)/m_0^2$ and kept terms up to order $1/\gamma^5$.
$A$ and $B$ are specified by Eq.~(\ref{bed}) and below.
Note that the effective Lagrangian~(\ref{leegavlag}) is valid up to ${\cal O}(\gamma^{-5})$.
For the self-consistent potentials, we obtain
\begin{align}
S=&\;m\left\{1+\frac{s_{22}}{\cosh^2 z}\frac{1}{\gamma^2}+\left( \frac{s_{42}}{\cosh^2 \xi}+\frac{s_{44}}{\cosh^4\xi}\right)\frac{1}{\gamma^4}\right. \nonumber\\
&{}+\left.\left( \frac{s_{52}}{\cosh^2 \xi}+\frac{s_{54}}{\cosh^4\xi}\right)\frac{1}{\gamma^5}\right. \label{ss}\\
&{}+\left.\left( \frac{s_{62}}{\cosh^2 \xi}+\frac{s_{64}}{\cosh^4\xi} + \frac{s_{66}}{\cosh^6 \xi}\right)\frac{1}{\gamma^6}{}+\cdots\right\}\nonumber
\end{align}
and
\begin{align}
V_0=&\;m\left\{\frac{\rho_{22}}{\cosh^2 z}\frac{1}{\gamma^2}+\left( \frac{\rho_{42}}{\cosh^2 \xi}+\frac{\rho_{44}}{\cosh^4\xi}\right)\frac{1}{\gamma^4}\right. \nonumber\\
&{}+\left.\left( \frac{\rho_{52}}{\cosh^2 \xi}+\frac{\rho_{54}}{\cosh^4\xi}\right)\frac{1}{\gamma^5}\right. \label{vv0}\\
&{}+\left.\left( \frac{\rho_{62}}{\cosh^2 \xi}+\frac{\rho_{64}}{\cosh^4\xi} + \frac{\rho_{66}}{\cosh^6 \xi}\right)\frac{1}{\gamma^6}{}+\cdots\right\}\,.\nonumber
\end{align}
The explicit expressions for the coefficients $s_{mn}$, $\rho_{mn}$ can be found in Appendix~\ref{app2}.
In this particular example, we have demonstrated that a consistent truncation schema based on an expansion parameter can be found.
As a result, we have obtained an analytically computable expression for $M_n$ in powers of $\gamma^{-1}$.
As pointed out above, exact bound-state solutions are not known for this model so that we have derived new results. They could in principle be checked numerically.

The analysis of localized bound states in 3+1 dimensions would presumably require numerical methods and is outside the scope of the present work.

In the following section we rather take a step back and discuss several more general features of the NSET approach in the context of the Walecka model. In particular we confront the NSET approach with the MFT approach with the no-sea approximation.

\section{General observations and results} \label{4}
\subsection{MFT with the no-sea approximation}

In order to clarify the differences, let us shortly review the MFT approach in the context of the Walecka model \cite{Serot:1984ey}.
MFT can be seen as an {\it ad hoc} prescription to gain practical solutions for the complicated field theoretic problem, posed by the Lagrangian~(\ref{Lag3+1}).
While the fermion field is still considered as a field operator, the meson fields are replaced by their expectation values, which are assumed to be classical fields with finite values.
The Dirac sea, occupied with an infinite number of negative-energy fermion states, is simply omitted. In result, there do not occur any divergences and the ``bare parameters" in Eq.~(\ref{Lag3+1}) are directly interpreted as their physical counterparts.
As above, we do not allow for physical couplings cubic and quartic in the scalar meson field and hence assume $\tilde{c}_3=\tilde{c}_4=0$.
In result, the Euler-Lagrange equation for $\psi$ becomes
\begin{equation}
 \left[{\rm i}\gamma_{\mu}\partial^{\mu}-g_{\omega}\gamma_{\mu}\langle\omega^{\mu}\rangle-(m-g_{\sigma}\langle\sigma\rangle)\right]\psi=0\,. \label{eq3b}
\end{equation}
The classical meson fields satisfy the field equations
\begin{align}
 ({\bf \nabla}^2-m_{\sigma}^2)\langle\sigma\rangle&=-g_{\sigma}\langle\bar\psi\psi\rangle\,, \label{eq1b}\\
 ({\bf \nabla}^2-m_{\omega}^2)\langle\omega^0\rangle&=-g_{\omega}\langle\psi^{\dag}\psi\rangle\,. \label{eq2b}
\end{align}
Since one considers matter at rest, the classical three vector field $\langle${\boldmath$\omega$}$\rangle$ is assumed to vanish \cite{Serot:1997xg}.
Hence, the classical meson fields are generated self-consistently by the fermion condensates.

We formally rewrite Eqs.~(\ref{eq1b}) and (\ref{eq2b}) as
\begin{align}
 \langle\sigma\rangle&=-g_{\sigma}\frac{1}{{\bf \nabla}^2-m_{\sigma}^2}\langle\bar\psi\psi\rangle\,, \nonumber\\
 \langle\omega^0\rangle&=-g_{\omega}\frac{1}{{\bf \nabla}^2-m_{\omega}^2}\langle\psi^{\dag}\psi\rangle\,, \label{conds}
\end{align}
and insert them in Eq.~(\ref{eq3b}). This yields the following equation of motion
\begin{align}
&\left[{\rm i}\gamma_{\mu}\partial^{\mu}+\gamma_0g_{\omega}^2\frac{1}{{\bf \nabla}^2-m_{\omega}^2}\langle\psi^{\dag}\psi\rangle\right. \nonumber\\
&\left.{}-\left(m+g_{\sigma}^2\frac{1}{{\bf \nabla}^2-m_{\sigma}^2}\langle\bar\psi\psi\rangle\right)\right]\psi=0\,.
\label{eqmotmft}
\end{align}

It is important to note that while the original field equations obviously refer to the full fermion field $\psi$, in MFT one uses them as equations for $\psi_+$ only, simply substituting $\psi$ for $\psi_+$. From a field theoretic point of view, such a drastic procedure seems hard to justify. However, the MFT approach turned out to describe many physical phenomena surprisingly well. 

\subsection{Uniform nuclear matter}

Let us first focus on (uniform) nuclear matter.
Keeping only the leading-order interaction terms in the NSET Lagrangian for spatially uniform matter, Eq.~(\ref{Lefftralainv}), we obtain Eq.~(\ref{leegavlag}).
The effective mass and the effective vector self-energy are then given by
\begin{align}
   M&=m-\frac{g_{\sigma}^2}{m_{\sigma}^2}\langle\bar\psi\psi\rangle\,, \nonumber\\
   V^{\nu}&=-\frac{g_{\omega}^2}{m_{\omega}^2}\langle\psi^{\dag}\psi\rangle\delta^{\nu}_0\,,
   \label{mmft}
\end{align}
and the associated equations of motion can be written as
\begin{equation}
\left[{\rm i}\slashed{\partial}-\left(m-\frac{g_{\sigma}^2}{m_{\sigma}^2}\langle\bar\psi\psi\rangle\right)-\gamma_{0}\frac{g_{\omega}^2}{m_{\omega}^2}\langle\psi^{\dag}\psi\rangle\right]\psi=0\,.
\label{eqmotmf}
\end{equation}
Note that Eqs.~(\ref{mmft}) and (\ref{eqmotmf}) are the defining equations of the MFT approach to stationary, uniform nuclear matter, where $\langle\sigma\rangle$ and $\langle\omega^0\rangle$ are assumed to be constants.
Let us, however, emphasize that the MFT approach significantly differs from the NSET approach.

In contrast to MFT, the NSET can be traced back to a relativistic Hartree approximation of the full, underlying relativistic quantum field theory, Eq.~(\ref{Lag3+1}). 
The Dirac sea is treated explicitly and enters the derivation of the effective theory in the form of $``-"$ loops. This gives rise to inevitable divergences, whose treatment requires an adequate renormalization procedure. In the course of this, renormalization conditions have to be specified. The resulting effective theory is dependent on these renormalization conditions. Their choice dictates the explicit form of the various couplings in the NSET. We choose them such that the effective quantities resemble their bare counterparts in the vicinity of the fully occupied Dirac sea.
Only then, having performed the lengthy derivation of the NSET, switching from bare to physical parameters, the correspondence noted above can be established.
In this respect the NSET approach indicates why (as long as a Hartree treatment is justified) the MFT approach yields reasonable results for uniform nuclear matter at small Fermi momentum in the Walecka model.

Note that the no-sea effective theory Lagrangian~(\ref{leegavlag}) should be valid up to ${\cal O}(k_f^{2})$ in 1+1 dimensions and up to ${\cal O}(k_f^{14})$ in 3+1 dimensions, the lowest-order correction term being $\sim(\bar\psi\psi)^3\sim k_f^3$ and $\sim(\bar\psi\psi)^5\sim k_f^{15}$, respectively (cf. the remarks in Sec.~\ref{3}).

It is easy to derive the energy of uniform matter at finite Fermi momentum within the framework of the NSET. One simply has to solve the single.particle equations of motion, derived from the respective effective Lagrangian ${\cal L}_{\rm eff}'$, for their energy eigenvalues and integrate them up. Note that in Hartree approximation this corresponds to integrating over the positive-energy solutions of the massive Dirac equation (mass $M$), featuring a chemical potential $V_{0}$. In addition one has to account for double counting corrections (dcc).
One obtains
\begin{align}
{\cal E}=&\;2N\int_0^{k_f}\frac{{\rm d}^D p}{(2\pi)^D} \left(E(p)-V_0\right) {}+ {\rm dcc}\,,
\label{energy}
\end{align} 
where $E(p)=\sqrt{{\bf p}^2+M^2}$, with $M$ from Eq.~(\ref{Mtrala}).
Specifying to 3+1 dimensions, Eq.~(\ref{energy}) becomes
\begin{align}
{\cal E}=&\;2N\int_0^{k_f}\frac{{\rm d}^3 p}{(2\pi)^3} E(p)+\frac{g_{\sigma}^2}{2m_{\sigma}^2}\langle\bar\psi\psi\rangle^2+\frac{g_{\omega}^2}{2m_{\omega}^2}\langle\psi^{\dag}\psi\rangle^2 \nonumber\\ &{}-\frac{2}{5m}\frac{N}{(2\pi)^2}\left(\frac{g_{\sigma}^2}{m_{\sigma}^2}\right)^5\langle\bar\psi\psi\rangle^5{}+\cdots\,.
\label{energy3+1}
\end{align} 
While the first three terms in Eq.~(\ref{energy3+1}) constitute the full MFT result, the last term written explicitly is the first nonvanishing NSET correction ($\sim k_f^{15}$). Since the lowest-order term omitted in Eq.~(\ref{energy3+1}) is proportional to $\langle\bar\psi\psi\rangle^6\sim k_f^{18}$, the terms written explicitly should reproduce the exact result up to ${\cal O}(k_f^{17})$. In this sense, the MFT result in 3+1 dimensions should be valid up to ${\cal O}(k_f^{14})$.

These predictions can indeed be verified by turning to the series expansion of the exact result for the energy of uniform matter, obtained by performing the full RHA calculation in 3+1 dimensions \cite{Serot:1984ey}
\begin{align}
{\cal E}_{\rm RHA}=&\;2N\int_0^{k_f}\frac{{\rm d}^3p}{(2\pi)^3}\sqrt{{\bf p}^2+M^2}{}+\frac{m_{\sigma}^2}{2g_{\sigma}^2}(M-m)^2 \nonumber\\
&{}+\frac{g_{\omega}^2}{2m_{\omega}^2}\rho_B^2-\frac{N}{2(2\pi)^2}\left\{M^4\ln\left(\frac{M}{m}\right)-\frac{25}{12}M^4\right.\nonumber\\
&{}+\left.4mM^3-3m^2M^2+\frac{4}{3}m^3M-\frac{1}{4}m^4\right\}\,.
\label{Eexact}
\end{align}

We now turn to the energy per nucleon, determined by ${\cal E}/\rho_B$. Obviously, the NSET result for ${\cal E}/\rho_B$ with ${\cal E}$ determined by Eq.~(\ref{energy3+1})
is valid up to ${\cal O}(k_f^{14})$ and can systematically be improved, while the MFT result can be trusted up to ${\cal O}(k_f^{11})$ only.

For illustration the series expansion of the NSET result up to ${\cal O}(k_f^{14})$ is stated explicitly
\begin{widetext}
\begin{align}
\frac{\cal E}{\rho_B}=m&+\left[\frac{3(k_f)^2}{10m}-\frac{3(k_f)^4}{56m^3}+\frac{(k_f)^6}{48m^5}-\frac{15(k_f)^8}{1408m^7}+\frac{21(k_f)^{10}}{3328m^9}-\frac{21(k_f)^{12}}{5120m^{11}}+\frac{99(k_f)^{14}}{34816m^{13}}{}+\cdots\right] \nonumber\\
&{}+\frac{g_{\omega}^2}{2m_{\omega}^2}\rho_B-\frac{g_{\sigma}^2}{2m_{\sigma}^2}\rho_B+\frac{g_{\sigma}^2}{m_{\sigma}^2}\frac{\rho_B}{m}\left[\frac{3(k_f)^2}{10m}-\frac{36(k_f)^4}{175m^3}+\frac{16(k_f)^6}{105m^5}-\frac{64(k_f)^8}{539m^7}+\frac{96(k_f)^{10}}{1001m^9}{}+\cdots\right] \nonumber\\
&{}+\left(\frac{g_{\sigma}^2\rho_B}{m_{\sigma}^2m}\right)^2\left[\frac{3(k_f)^2}{10m}-\frac{351(k_f)^4}{700m^3}+\frac{8803(k_f)^6}{14000m^5}{}-\frac{305923(k_f)^8}{431200m^7}{}+\cdots\right] \nonumber\\
&{}+\left(\frac{g_{\sigma}^2\rho_B}{m_{\sigma}^2m}\right)^3\left[\frac{3(k_f)^2}{10m}-\frac{69(k_f)^4}{70m^3}{}+\cdots\right]+\left(\frac{g_{\sigma}^2\rho_B}{m_{\sigma}^2m}\right)^4\left[\frac{g_{\sigma}^2}{m_{\sigma}^2}\frac{Nm^3}{40\pi^2}+\frac{3(k_f)^2}{10m}\left(1-\frac{g_{\sigma}^2}{m_{\sigma}^2}\frac{Nm^2}{8\pi^2}\right){}+\cdots\right].
\label{energie3+1durchrho}
\end{align}
\end{widetext}
Here, we take over the grouping, as well as the interpretation of the different contributions from \cite{Serot:1997xg}, Eq.~(4.8).
The first term is the fermion rest mass, followed by the nonrelativistic Fermi-gas energy and the first few relativistic corrections. The next two terms (proportional to $\rho_B$) give the nonrelativistic limit of the potential energy coming from the vector and scalar mesons.
The following term in brackets (with overall factor $\rho_B$) is a relativistic correction to the scalar potential energy that arises from the Lorentz contraction factor in the scalar density, evaluated for fermions of mass $m$. The next three terms (with overall factors $\rho_B^2$, $\rho_B^3$ and $\rho_B^4$) are also corrections to the scalar potential energy.

Let us shortly discuss the physical range of $k_f$, where such an expansion is sensible. Turning to the dimensionless quantity $\frac{1}{m}\frac{\cal E}{\rho_B}$, $k_f$ appears in the dimensionless ratios $k_f/m$, $k_f/m_{\sigma}$, and $k_f/m_{\omega}$ only. The smallest mass is governing the range of validity of the expansion. With $m=939\,{\rm MeV}$, $m_{\sigma}=550\,{\rm MeV}$, and $m_{\omega}=783\,{\rm MeV}$ \cite{Serot:1984ey}, the smallest mass corresponds to $m_{\sigma}$.
Assuming the ratio to be small when it is $\leq 0.5$, we obtain the following estimate $k_f/m_{\sigma}=k_f/(2.8\,{\rm fm}^{-1})\leq 0.5$. Hence, we expect a good convergence for $k_f\leq 1.4\,{\rm fm}^{-1}$ Note that this is in agreement with the convergence behavior of the curves depicted in Fig.~\ref{energykf14} (see below).

The series expansions of the full MFT and RHA results for the energy per nucleon in uniform nuclear matter agree with each other up to ${\cal O}(k_f^{11})$. By including higher-order interaction terms in the NSET Lagrangian, the NSET result can be systematically improved to reproduce the RHA result up to any desired order. The higher-order terms constituting the full MFT result cannot be justified in the NSET approach.

\begin{figure}
	\begin{center}
	\epsfig{file=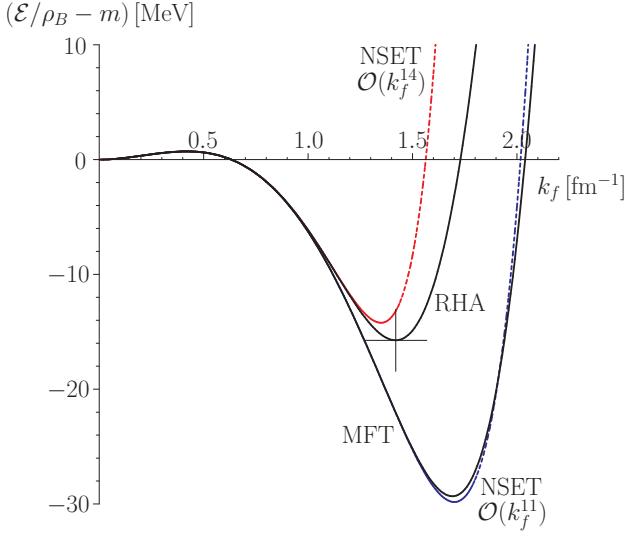,width=8.2cm,angle=0}
	\end{center}
	\caption{The energy per nucleon in uniform nuclear matter with $N=2$ in 3+1 dimensions determined in full MFT and RHA, respectively, is confronted with the NSET results valid up to ${\cal O}(k_f^{11})$ and ${\cal O}(k_f^{14})$. The coupling constants are chosen to fit the value and position of the minimum of the RHA curve to the phenomenological values $k_f^0=1.42\,{\rm fm}^{-1}$, $({\cal E}/\rho_B-m)=-15.75\,{\rm MeV}$ denoted by the cross symbol. With $m=939\,{\rm MeV}$, $m_{\sigma}=550\,{\rm MeV}$, $m_{\omega}=783\,{\rm MeV}$, we obtain $g^2_{\sigma}=62.89$ and $g^2_{\omega}=79.78$.}
	\label{energykf14}
\end{figure}
To clarify the situation, we consider an illustration (cf. Fig.~\ref{energykf14}) as helpful. However, several comments are in order here. Obviously, a plot of the energy per nucleon requires a choice of the coupling constants. Trying to describe the physics by MFT and RHA, respectively, one fixes the coupling constants differently in both approaches. Namely in that way that the minimum of the corresponding curve coincides with the phenomenological values $k_f^0=1.42\,{\rm fm}^{-1}$, $({\cal E}/\rho_B-m)=-15.75\,{\rm MeV}$ \cite{Serot:1984ey}. Here, however, our intention is different. Considering the Walecka model as underlying relativistic QFT, we see the MFT as ad-hoc description, simply omitting the Dirac sea. The RHA, accounting for the Dirac sea, is fitted to the phenomenological values of the minimum. Thereby the coupling constants are fixed and used for all curves (RHA, MFT, NSET). Hence, Fig.~\ref{energykf14} rather shows, how the NSET curve approaches the RHA curve by including higher-order terms in an series expansion in $k_f$.

For small Fermi momenta $k_f\lesssim1.1\,{\rm fm}^{-1}$ all the curves agree nicely but start to deviate for larger values of $k_f$.
While the MFT and RHA curves share the same general characteristics, namely a local minimum and a steeply rising behavior toward larger Fermi momenta, the positions of the minima differ by a Fermi momentum of $\cong 0.27\,{\rm fm}^{-1}$ and an energy $\cong 14\,{\rm MeV}$. Including the first NSET correction term $\sim(\bar\psi\psi)^5$ in Eq.~(\ref{energy3+1}), the corresponding curve valid up to ${\cal O}(k_f^{14})$ clearly approaches the RHA curve.

\subsection{Spatially nonuniform systems}

Let us extend the analysis of uniform nuclear matter in its ground state to include spatially inhomogeneous systems.
This is necessary to describe finite ``nuclei" or bound states.

One should note that it has been shown \cite{Friman:1988ma,Price:1990dj,Price:1990zz} within the Walecka model, that the ground state of infinite nuclear matter at a given finite Fermi momentum $k_f$ does not necessarily imply the fermion condensates to be translationally invariant. However, this is not in conflict with the NSET approach, which is based only on the assumption of translational invariance of the vacuum ($k_f=0$). By accounting for spatial inhomogeneous condensates in the ``+'' sector, we rather explicitly allow for the formation of such a translationally noninvariant state at finite $k_f$.

While the MFT approach to stationary, spatially uniform matter can be considered as leading-order treatment within the NSET approach, the situation is different for spatially nonuniform systems. Here, the respective MFT equations cannot be inferred directly from the NSET Lagrangian.
In contrast to our NSET Lagrangian, the assumption of an almost local effective theory, explicitly restricting to small momentum transfer, is not yet implemented in Eq.~(\ref{eqmotmft}).
It is instructive to recall Eq.~(\ref{Leff2int}) and note that the scalar self-energy can be written as
\begin{equation}
S(k)=m+g_{\sigma}^2\Delta'(k)\langle\bar\psi\psi\rangle_k{}+\cdots\,.
\end{equation}
Here the ellipses denote further contributions, which are of higher powers in the fermion condensates.
The vector part is determined analogously
\begin{equation}
 V_{\nu}(k)=-g_{\omega}^2 D'^{\ \mu}_{\nu}(k)\langle\bar\psi\gamma_{\mu}\psi\rangle_k{}+\cdots\,.
 \label{signu}
\end{equation}

\begin{figure}
	\begin{center}
	\epsfig{file=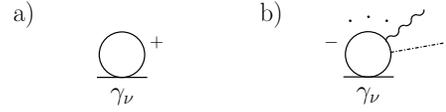,width=5.8cm,angle=0}
	\end{center}
	\caption{Illustration of the effective fermion current $\langle\bar\psi\gamma_{\nu}\psi\rangle$. a) Valence part $\langle\bar\psi\gamma_{\nu}\psi\rangle_+$. b) Residual contribution to the fermion current starting with a $``-"$ loop. It is given by the infinite sum of topologically distinct diagrams, made up of further internal $``-"$ loops and outermost $``+"$ loops, coupled via $\sigma$ and $\omega$ propagators.}
	\label{fermcurrent}
\end{figure}
In the Walecka model the fermion current is conserved
\begin{equation}
\partial^{\nu}\langle\bar\psi\gamma_{\nu}\psi\rangle=0 \quad \leftrightarrow \quad k^{\nu}\langle\bar\psi\gamma_{\nu}\psi\rangle_k=0\,.
\label{curcon1}
\end{equation}
Note that $\psi$ in Eqs.~(\ref{curcon1}) is the full fermion field. In contrast to uniform nuclear matter, it is not clear {\it a priori} that an analogous relation for $``+"$ states holds for spatially nonuniform systems also. The expression for the fermion current translates into an infinite sum of topologically distinct diagrams (cf. Fig.~\ref{fermcurrent}).
Turning to Eq.~(\ref{curcon1}), the momentum $k^{\nu}$ enters the initial $``-"$ loop, Fig.~\ref{fermcurrent}b, in the form of $k^{\nu}\gamma_{\nu}=\slashed{k}$. It can be shown that the contribution of a given $``-"$ loop, with arbitrary $1$ and $\gamma^{\mu}$ insertions, vanishes if one inserts an additional momentum via $\slashed{k}$. To account for all topologically distinct diagrams, one has to sum over all possible insertion points. 

As a result one obtains
\begin{equation}
\partial^{\nu}\langle\bar\psi\gamma_{\nu}\psi\rangle_+=0\,.
\label{curcon3}
\end{equation}
For stationary systems, Eq.~(\ref{curcon3}) becomes $\nabla\langle\bar\psi{\bm\gamma}\psi\rangle_+=0$.
This is in particular fulfilled for $\langle\bar\psi{\bm\gamma}\psi\rangle_+=0$, corresponding to ``positive-energy matter" at rest, which we will consider in the remainder of this section.

Transforming into position space and inserting the explicit expressions for the effective meson propagators, assuming stationary condensates, we obtain
\begin{equation}
 S(x)=m+g_{\sigma}^2\left(\frac{\rm 1}{{\bf \nabla}^2-m_{\sigma}^2}+\mathcal{O}\left(({\bf \nabla}^2)^2\right)\right)\!\langle\bar\psi\psi\rangle{}+\,\cdots
 \label{Mx}
\end{equation}
and
\begin{equation}
 V^{0}(x)=g_{\omega}^2\left( \frac{1}{{\bf \nabla}^2-m_{\omega}^2}+\mathcal{O}\left(({\bf \nabla}^2)^2\right)\right)\!\langle\psi^{\dag}\psi\rangle{}+\,\cdots\,,
 \label{Sigmax}
\end{equation}
which enter the equation of motion for the $``+"$ fermion field
\begin{equation}
 \left({\rm i}\slashed{\partial}+\gamma_0V^0(x)-S(x)\right)\psi=0\,.
 \label{eqmotx}
\end{equation}

Concentrating on Eqs.~(\ref{eqmotmft}) and (\ref{eqmotx}), MFT dealing with stationary, nonuniform systems and the NSET approach can be compared.

The MFT result can be recovered from the NSET approach by omitting both the $\mathcal{O}\left(({\bf \nabla}^2)^2\right)$ terms in Eqs.~(\ref{Mx}) and (\ref{Sigmax}), as well as the Dirac sea-induced terms of higher loop order, involving increasing powers in the fermion condensates and effective propagators, indicated by the ellipses.
In the framework of the NSET, we do not see how such an ``approximation" could be justified.

Let us finally turn to the special case where the meson masses dominate the denominators of the meson propagators.
This was assumption in the construction of our NSET Lagrangians. Only then do the interactions become almost local and effective coupling constants can be defined.
The equation replacing Eq.~(\ref{eqmotx}) is
\begin{align}
&\left\{{\rm i}\slashed{\partial}-\frac{g_{\omega}^2}{m_{\omega}^2}\gamma_0\left(1+\frac{\nabla^2}{m_{\omega}^2}+c_{024}\frac{(\nabla^2)^2}{m_{\omega}^2}{}+\cdots\right)\langle\psi^{\dag}\psi\rangle\right. \nonumber\\
&-\left.m+\frac{g_{\sigma}^2}{m_{\sigma}^2}\left(1+\frac{\nabla^2}{m_{\sigma}^2}+c_{204}\frac{(\nabla^2)^2}{m_{\sigma}^2}{}+\cdots\right)\langle\bar\psi\psi\rangle\right. \label{eqmotx2}\\
&\hspace*{5.15cm}\left.\phantom{\frac{\nabla^2}{m_{\omega}^2}}+\cdots\right\}\psi=0\,. \nonumber
\end{align}
The ellipses in the last line denote contributions that are of higher powers in the fermion condensates, descending from effective interactions not present in the original Lagrangian.
Equation~(\ref{eqmotmft}) becomes
\begin{align}
&\left\{{\rm i}\slashed{\partial}-\frac{g_{\omega}^2}{m_{\omega}^2}\gamma_0\left(1+\frac{\nabla^2}{m_{\omega}^2}+\frac{(\nabla^2)^2}{m_{\omega}^4}{}+\cdots\right)\langle\psi^{\dag}\psi\rangle\right. \label{eqmotmft2} \\
&-\left.m+\frac{g_{\sigma}^2}{m_{\sigma}^2}\left(1+\frac{\nabla^2}{m_{\sigma}^2}+\frac{(\nabla^2)^2}{m_{\sigma}^4}{}+\cdots\right)\langle\bar\psi\psi\rangle
\right\}\psi=0\,. \nonumber
\end{align}
Keeping terms up to ${\cal O}(\nabla^2)$ in the expansion of the meson propagators only, Eqs.~(\ref{eqmotx2}) and (\ref{eqmotmft2}) coincide.
This is a consequence of our renormalization procedure.
We chose $\alpha_2$, $\alpha_{\sigma}$ and $\alpha_{\omega}$ such that up to this order the effective meson propagators in the NSET resemble their counterparts in the underlying relativistic field theory. 
Turning to higher-order terms, differences are unveiled.

The coefficients of the $(\nabla^2)^2$ derivatives of the condensates differ by contributions proportional to the square of the coupling constants.
Moreover, the effective multi-fermion interactions, arising in the construction of the NSET, are not accounted for within the MFT.

Hence, in contrast to the MFT approach to uniform matter, in general the MFT approach to nonuniform systems cannot be seen as a systematic leading-order treatment within the framework of the NSET.
We believe that the NSET approach provides us with a new tool to account for the effects of the Dirac sea in a controllable manner, avoiding {\it ad hoc} approximations.
Applications of this method to spatially nonuniform systems will require numerical studies and have not yet been performed.

\section{Summary and Conclusions}\label{5}
In this article, we have constructed a no-sea effective theory for the Walecka model. We focused on the relativistic Hartree approximation. While the Dirac sea is integrated out, the positive-energy fermion states are treated explicitly. The resulting NSET Lagrangian features an infinite number of effective interaction terms. For the case of uniform nuclear matter the NSET could be tested quantitatively. Restricting to 1+1 dimensions, new analytical insights in uniform nuclear matter were obtained.

Finally, we confronted our NSET approach with the standard MFT approach with the no-sea approximation to the Walecka model, wherein the Dirac sea is simply omitted. It turned out that MFT applied to uniform nuclear matter at small Fermi momentum can be reinterpreted as leading-order treatment of the NSET.
We have determined the order in a series expansion in $k_f$ to which the MFT results can be trusted; e.g., for the 3+1 dimensional Walecka model, the energy per nucleon determined by the MFT approach is valid only up to ${\cal O}(k_f^{11})$. Within the NSET approach, the order of validity can be improved systematically.
Extending the analysis to spatially nonuniform systems, the situation turned out to be more complicated.
Here we emphasized the significant differences between the two approaches. 

\begin{acknowledgments}
I thank Michael Thies for many interesting discussions and helpful remarks.
\end{acknowledgments}

\appendix


\section{Exact localized solutions of no-sea effective theory with four-fermion-interactions}
\label{app1}

In this section we use an approach, introduced by Lee {\it et al}. \cite{Lee:1975tx} to construct exact localized ``classical" solutions of 1+1 dimensional field theories with four-fermion-interactions. To apply their considerations to Eq.~(\ref{leegavlag}), we have to generalize their ideas to the simultaneous appearance of scalar and vector interactions featuring two different coupling constants $G_{\sigma}\neq G_{\omega}$,
\begin{align}
  \mathcal{L}=&\;\bar\psi\left({\rm i}\slashed{\partial}-m\right)\psi\,+\,\frac{G_{\sigma}^2}{2}\left(\bar\psi\psi\right)^2 \nonumber\\
  &{}+\,\frac{G_{\omega}^2}{2}\left(\bar\psi{\rm i}\gamma_{\nu}\psi\right)\left(\bar\psi{\rm i}\gamma^{\nu}\psi\right) .
  \label{Lag3eff1}
\end{align}
Here $\psi$ corresponds to a $``+"$ fermion spinor with $N$ flavors.
The required generalization is straight-forward.
Following Lee {\it et al}., fermion bound states can be constructed. To retain analyticity, they assume that all fermions occupy one and the same energy level (energy eigenvalue $E$). Obviously, this limits their approach to fermion bound states with $n\leq N$ fermions only.
It is convenient to turn to the notation of \cite{Lee:1975tx}. This implies the following choice of the $\gamma$ matrices
\begin{equation}
  \gamma_0=\sigma_3\,, \quad \gamma_1=-{\rm i}\sigma_1\,, \quad \gamma_5=\gamma_0\gamma_1=\sigma_2\,,
  \label{gammam}
\end{equation}
and the ansatz ($k$ denotes the flavor index) 
\begin{equation}
  \psi_k(x)=\binom{u_k}{v_k}=\frac{R(x)}{\sqrt{N}}\binom{\cos\theta}{\sin\theta} ,
  \label{psi}
\end{equation}
where $u_k$, $v_k$ have to be real valued functions (\cite{Lee:1975tx}, below Eq.~(7c)). $\theta(x)$ is determined by
\begin{equation}
  \theta(x)={\rm arctan}\left(\alpha\,{\rm tanh}\beta x\right) .
  \label{theta}
\end{equation}
$\alpha$ and $\beta$ are energy- and mass-dependent constants defined by
\begin{equation}
  \alpha=\sqrt{\frac{m-E}{m+E}} \quad {\rm and} \quad \beta=\sqrt{m^2-E^2}\,.
  \label{alphabeta}
\end{equation}
By using Ref.~\cite{Lee:1975tx}, Eq.~(8a), together with the Lagrangian~(\ref{Lag3eff1}), we infer
\begin{equation}
  R^2(x)=\frac{2\left(E-m\cos{2\theta}\right)}{G_{\omega}^2-G_{\sigma}^2\,\cos^2{2\theta}}\,.
  \label{RR}
\end{equation}
Demanding the following normalization
\begin{equation}
  n=\int\limits_{-\infty}^{\infty}{\rm d}x\,\psi^{\dag}\psi=\int\limits_{-\arctan\alpha}^{\arctan\alpha}{\rm d}\theta\,\frac{-2}{G_{\omega}^2-G_{\sigma}^2\,\cos^2{2\theta}}\,,
  \label{intRR=n}
\end{equation}
where $n\geq0$,
Eq.~(\ref{intRR=n}) can be fulfilled consistently only for $G_{\sigma}^2>G_{\omega}^2$. In this case we obtain
\begin{equation}
  2\,\arctan\alpha=\arctan\left[\frac{G}{G_{\omega}}\tanh\left(\frac{n}{2}G_{\omega}G\right)\right],
  \label{intRR=n2}
\end{equation}
where we defined
\begin{equation}
G=\sqrt{G_{\sigma}^2-G_{\omega}^2}\,.
\end{equation}
Equation~(\ref{intRR=n2}) can be also written as
\begin{equation}
  \frac{\alpha}{1-\alpha^2}=\frac{G}{2\,G_{\omega}}\tanh\left(\frac{n}{2}G_{\omega}G\right) .
  \label{intRR=n3}
\end{equation}
The ``bound-state mass" of $n$ fermions corresponds to the expectation value of the classical Hamiltonian. It is given by (cf. Ref.~\cite{Lee:1975tx}, Eq. (7c))
\begin{align}
  M_n&=m\int\limits_{-\infty}^{\infty}{\rm d}x\,\bar\psi\psi=m\int\limits_{-\arctan\alpha}^{\arctan\alpha}{\rm d}\theta\,\frac{-2\cos{2\theta}}{G_{\omega}^2-G_{\sigma}^2\,\cos^2{2\theta}} \nonumber\\
  &=\frac{2m}{G_{\sigma}G}\,{\rm arctanh}\left[\frac{G_{\sigma}}{G}\sin\left(2\,\arctan\alpha\right)\right] \label{zw} \\
  &=\frac{2m}{G_{\sigma}G}\,{\rm arcsinh}\left[\frac{G_{\sigma}}{G_{\omega}}\sinh\left(\frac{n}{2}G_{\omega}G\right)\right] .
  \label{M}
\end{align}
In the last step of Eq.~(\ref{M}) we inserted Eq.~(\ref{intRR=n2}).
Let us investigate the stability of the $n$-fermion localized solution.
To be stable, it has to be energetically favored as compared to the mass of $n$ free fermions, as well as the mass of $n$ localized fermions not interacting with each other.
The first condition is true, as
\begin{equation}
   {\rm Arsinh}(w\sinh z)\leq wz
\end{equation}
for $w\geq 1\,, z\geq0$ directly implies $M_n\leq nm$. To fulfill the latter condition, we have to ensure that $M_n\leq nM_1$. This is also true as
\begin{equation}
   {\rm Arsinh}(w\sinh(nz))\leq n\,{\rm Arsinh}(w\sinh z)
\end{equation}
for $w\geq 1\,, z\geq0$.
In result, the $n$-fermion localized solution with mass $M_n$ is energetically favored.
Finally, we determine the energy eigenvalue $E$. It is given by
\begin{equation}
  E=m\,\frac{1-\alpha^2}{1+\alpha^2}\,.
  \label{E}
\end{equation}
Recasting Eq.~(\ref{zw}) into
\begin{align}
  \frac{\alpha}{1+\alpha^2}&=\frac{1}{2}\sin\left(2\,\arctan\alpha\right) \nonumber\\
  &=\frac{G}{2G_{\sigma}}\tanh\left(\frac{G_{\sigma}G}{2m}M_n\right)
  \label{bed2}
\end{align}
and using Eq.~(\ref{intRR=n3}), one obtains
\begin{equation}
  E=m\,\frac{\cosh\left(\frac{n}{2}G_{\omega}\sqrt{G_{\sigma}^2-G_{\omega}^2}\right)}{\sqrt{1+\left(\frac{G_{\sigma}}{G_{\omega}}\right)^2\sinh^2\left(\frac{n}{2}G_{\omega}\sqrt{G_{\sigma}^2-G_{\omega}^2}\right)}}\,.
  \label{E1}
\end{equation}
Therewith, the explicit expression for the $n$-fermion-state spinor becomes
\begin{equation}
  \psi(x)=\frac{R(x)}{\sqrt{1+\alpha^2\tanh^2{\beta x}}}\binom{1}{\alpha\tanh{\beta x}}\,,
  \label{psi2}
\end{equation}
where we insert $\alpha$ and $\beta$ defined in Eq.~(\ref{alphabeta}) with $E$ from Eq.~(\ref{E1}). Note again that to be a solution, $\psi(x)$ has to be real valued.
Adopting the explicit values for the coupling constants, we obtain a solution of the Lagrangian~(\ref{leegavlag}), we are interested in here. 
For the special case $G_{\sigma}=G_{\omega}$ we refer to Ref.~\cite{Lee:1975tx}.

\section{Results for coefficients defined in Sec.~\ref{3a}}
\label{app2}

Here we collect the coefficients associated with the $n$-fermion bound state with mass $M_n$ in Sec.~\ref{3a}.
In order to simplify the notation, we define
\begin{equation}
\tilde{A}=\frac{Ag_{\sigma}^2}{m_0^2}\,, \quad \tilde{B}=\frac{Bg_{\omega}^2}{m_0^2}\,, \quad c=\tilde{A}-\tilde{B}\,.
\end{equation} 
Coefficients for the scalar potential $S(x)$, Eq.~(\ref{ss}), 
\begin{align}
s_{22}&=-\frac{n^2\tilde{A}c}{4}\,,\quad s_{42}=\frac{n^4\tilde{A}c^2}{96}(3\tilde{A}-5\tilde{B})\,, \nonumber\\
s_{44}&=\frac{n^4\tilde{A}c^2}{64}(\tilde{A}+3\tilde{B})\,, \nonumber\\
s_{52}&=-\frac{n^4\tilde{A}c}{72}\left[\frac{N\tilde{A}^3}{\pi}+\frac{6m^2c}{m_0^2}\left(4A\tilde{A}-3A\tilde{B}-B\tilde{B}\right)\right], \nonumber\\
s_{54}&=\frac{n^4\tilde{A}c}{96}\left[\frac{N\tilde{A}^2}{\pi}(3\tilde{B}-5\tilde{A})\right. \nonumber\\
&\hspace*{1cm}\left.\phantom{\frac{N\tilde{A}^2}{\pi}}{}+\frac{12m^2c}{m_0^2}\left(5A\tilde{A}-3A\tilde{B}-2B\tilde{B}\right)\right], \nonumber
\end{align}
\begin{align}
s_{62}&=-\frac{n^6\tilde{A}c^3}{23040}\left(135\tilde{A}^2-210\tilde{A}\tilde{B}+91\tilde{B}^2\right), \nonumber\\
s_{64}&=-\frac{n^6\tilde{A}c^3}{3072}\left(9\tilde{A}^2-14\tilde{A}\tilde{B}-27\tilde{B}^2\right), \nonumber\\
s_{66}&=-\frac{n^6\tilde{A}c^3}{1024}\left(\tilde{A}^2+10\tilde{A}\tilde{B}+5\tilde{B}^2\right).
\end{align}
\\
\\
Coefficients for the potential $V_0(x)$, Eq.~(\ref{vv0}),
\begin{align}
\rho_{22}&=-\frac{n^2\tilde{B}c}{4}\,, \nonumber\\
\rho_{42}&=-\frac{n^4\tilde{B}^2c^2}{48}\,, \nonumber\\
\rho_{44}&=\frac{n^4\tilde{B}c^2}{64}(3\tilde{A}+\tilde{B})\,, \nonumber\\
\rho_{52}&=-\frac{n^4\tilde{B}c}{72}\left[\frac{N\tilde{A}^3}{\pi}+\frac{6m^2c}{m_0^2}\left(A\tilde{A}+3\tilde{A}B-4B\tilde{B}\right)\right], \nonumber\\
\rho_{54}&=\frac{n^4\tilde{B}c}{48}\left[-\frac{N\tilde{A}^3}{\pi}+\frac{6m^2c}{m_0^2}\left(2A\tilde{A}+3\tilde{A}B-5B\tilde{B}\right)\right], \nonumber\\
\rho_{62}&=-\frac{n^6\tilde{B}^3c^3}{1440}\,, \nonumber\\
\rho_{64}&=-\frac{n^6\tilde{B}c^3}{3072}\left(15\tilde{A}^2-42\tilde{A}\tilde{B}-5\tilde{B}^2\right), \nonumber\\
\rho_{66}&=-\frac{n^6\tilde{B}c^3}{1024}\left(5\tilde{A}^2+10\tilde{A}\tilde{B}+\tilde{B}^2\right).
\end{align}

\bibliographystyle{abbrv}

\end{document}